\documentclass[fleqn,usenatbib]{mnras}

\usepackage{newtxtext,newtxmath}

\usepackage[T1]{fontenc}
\usepackage{ae,aecompl}
\usepackage{graphicx}
\usepackage{epstopdf}
\usepackage{xspace} 
\usepackage{natbib}

\bibpunct{(}{)}{;}{a}{}{,}

\newcommand{\kms}{\rm km\,s^{-1}}
\newcommand{\pc}{{\rm pc}}

\newcommand{\twomass}{\textit{2MASS}\xspace}

\usepackage{graphicx}	
\usepackage{amsmath}	
\usepackage{amssymb}	
\usepackage{gensymb}
\usepackage{booktabs,caption}
\usepackage[flushleft]{threeparttable}
\usepackage{mathtools}
\usepackage[dvipsnames]{xcolor}

\title[The link between magnetic fields and filamentary clouds II]{The link between magnetic fields and filamentary clouds II: Bimodal linear mass distributions}

\author[Law,C.-Y., et al.]{
Law, C.Y.,$^{1}$\thanks{E-mail: lawchiyan@hotmail.com}
Li, H. -B.,$^{1}$\thanks{E-mail: hbli@phy.cuhk.edu.hk}
and Leung, P.K.$^{1}$
\\
$^{1}$Department of Physics, The Chinese University of Hong Kong, Ma Liu Shui, Shatin, NT, Hong Kong\\
}

\date{Accepted XXX. Received YYY; in original form ZZZ}

\pubyear{2018}

\begin{document}
\label{firstpage}
\pagerange{\pageref{firstpage}--\pageref{lastpage}}
\maketitle

\begin{abstract}
By comparing cumulative linear mass profiles of 12 Gould Belt molecular clouds within $500\;{\rm \pc}$, we study how the linear mass distributions of molecular clouds vary with the angles between the molecular cloud long axes and the directions of the local magnetic fields (cloud-field direction offsets). We find that molecular clouds with the long axes perpendicular to the magnetic field directions show more even distributions of the linear mass. The result supports that magnetic field orientations can affect the fragmentation of molecular clouds (Li et al. 2017).
\end{abstract}




\section{Introduction}

The advance in both instrumentation and numerical simulation techniques in the past decade have unfolded the importance of magnetic fields in regulating clouds formation, evolution, and star-forming processes in molecular clouds  \citep{2012ARA&A..50...29C,2014prpl.conf..101L,10.1088/978-1-6817-4293-9}.

\begin{figure}
	\includegraphics[width=\columnwidth]{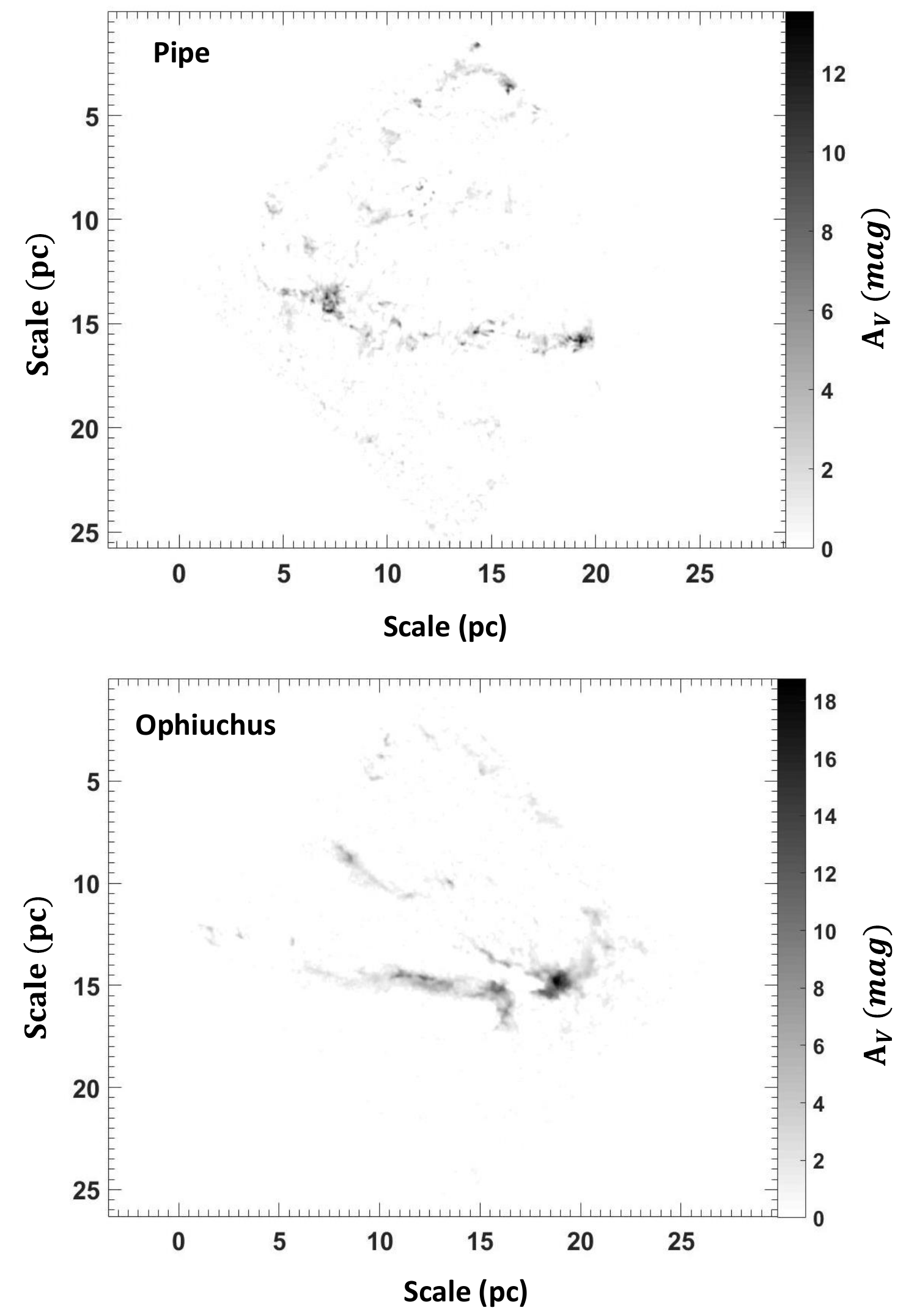}
    \caption{Dust extinction maps of Pipe and Ophiuchus. The extinction maps are rotated to align the long axis of the cloud to the horizontal axis. Mass accumulates toward one side of the filament of Ophiuchus, while Pipe has more balanced mass distribution along the long axis.} 
    \label{fig:Figure1}
\end{figure}

\cite{2013MNRAS.436.3707L} (hereafter paper-I) studied 13 Gould Belt molecular clouds and revealed that molecular cloud long axes tend to be either perpendicular (hereafter perpendicular alignment) or parallel (hereafter parallel alignment) to the mean directions of local magnetic fields. The bimodal distribution may affect the cloud fragmentation and star formation rate, as supported by both recent simulation and observational studies \citep[hereafter L17]{2015MNRAS.452.2410S,2017NatAs...1E.158L}. For example, inspection of the overall cloud morphologies of Ophiuchus and Pipe (Figure ~\ref{fig:Figure1} ), two neighboring clouds with significantly different star formation rates, suggests distinct spatial mass distributions. Ophiuchus with parallel alignment has mass clustered toward one end of the cloud (``head-heavy'' morphology), while Pipe with perpendicular cloud alignment shows relatively balanced mass distribution along the major cloud axis. The goal of this sequel is to systematically study whether all the perpendicular clouds result in a more even mass distribution compared to the parallel ones. As the first step, we design the test in the simplest possible way: comparing the total mass in the two halves of an elongated cloud.

For the goal to compare the two halves, we first need to define the volume of a cloud, which is introduced in the next section. Section 3 presents the statistical analysis of the results and discusses on the main findings. Section 4 is a summary.

\section{Method}
This work aims to study mass corresponding to star-forming activities. Simulations of isothermal turbulence with self-gravity show that the column density probability density function (N-PDF) will develop a power-law tail due to regions undergoing gravitational contractions \citep {2013ApJ...763...51F,2014ApJ...781...91G,2014MNRAS.445.1575W}. Such a regime is above a transition column density, at which the N-PDF starts to deviate from log-normal to power-law-like. This feature is indeed observed for molecular clouds with active star formation  \citep{2009A&A...508L..35K,2010MNRAS.406.1350F,2015A&A...577L...6S}. Paper-I also showed that the transition density of the N-PDFs from \cite{2009A&A...508L..35K} is very close to the magnetic critical column density,  which is the column density required to overcome magnetic pressure for gravitational contraction. Thus, we consider column densities above the transition density of the N-PDFs of molecular clouds. We identify the transition density by probing the changes in the slope of the N-PDF. Details on identifying the transition density are presented in Appendix A.   

To investigate the evenness of the mass distribution in the molecular clouds, we spatially divide a cloud into two halves and study the mass ratio between the two halves. This inevitably requires a well-defined cloud range to define a middle point. Since only the total mass in each half matters, we can simplify a map into a linear mass profile along the long axis of an elongated cloud.

We adopt the cloud long axis orientations and maps from L17. Recent studies suggest that the Perseus cloud is composed of two superimposed clouds with very different distances. \citep{1990ApJ...359..363G,2014ApJ...786...29S}. Thus, we discard the Perseus cloud from the analysis. Table 1 summaries the general information of the clouds. Here we summarise the  procedures of defining a cloud volume and its two halves. CrA serves as an example to illustrate the methods. 

\begin{figure*}
	\rotatebox[origin=c]{-90}{\includegraphics[width=14.5cm,height=19cm]{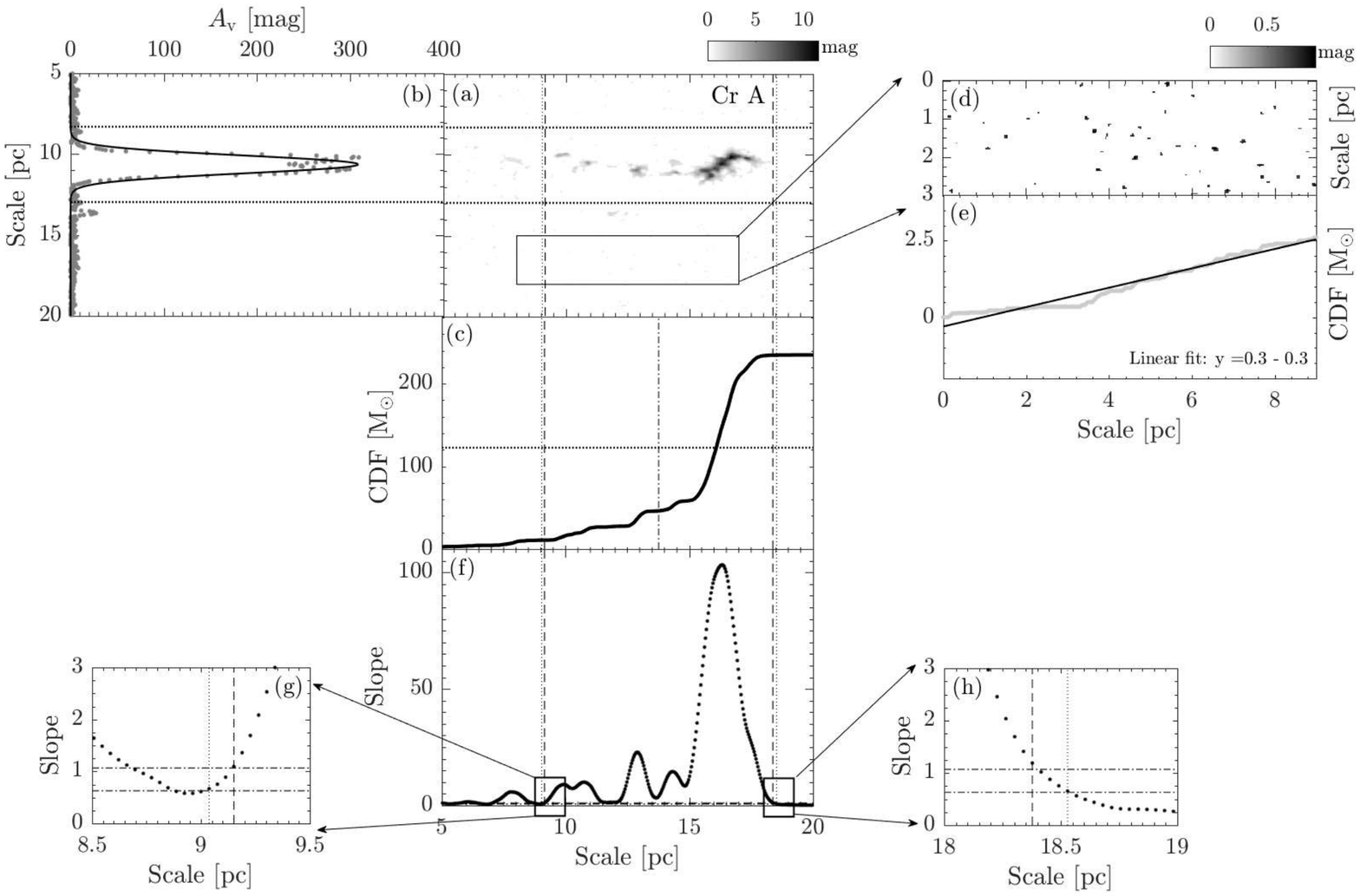}}
 \caption{ {\bf Panel a} : The extinction map of CrA. The map is rotated clockwise by the angle of the long-axis orientation from L17. The box indicates the reference region used to compute ${\rm slope}_{\rm ref}$, which is defined in panel e. For molecular clouds have distance less than 250pc (CrA,Ophiuchus, Pipe, Taurus, Musca, Lupus-I, Lupus-II-VI, and Cha-I-III), the dimension of the reference region is $9{\rm pc} \times 3{\rm pc}$. For molecular clouds have distance further than 250pc (Aquila, IC5146, Orion A, and Orion B), the dimension is enlarged to $15{\rm pc} \times 5{\rm pc}$. {\bf Panel b} : The gray dotted line presents the linear mass along the short axis. The overlaid solid line represents the Gaussian fit. The two horizontal black dotted lines are used to define the cloud width, with each line is 3-sigma from the peak. They are extended to panel a to show the boundaries of the cloud width. {\bf Panel c} : CLM profile of the extinction map within the cloud width. {\bf Panel d \& e} : The extinction map and CLM profile of the reference region for CrA. The solid line is the linear fit of the CLM profile. The corresponding linear function is presented in the lower right corner. We take the fitted slope as ${\rm slope}_{\rm ref}$. {\bf Panel f} : The dotted line presents the slope of CLM profile of the extinction map within the cloud width. Between the two vertical dashed lines, the slope is consistently 3 sigma above the mean of the ${\rm slope}_{\rm ref}$ from all the clouds in our study(figures~\ref{fig:Figure2} -~\ref{fig:Figure16}). The half-length of the cloud is thus defined by the middle point between the two lines, which is labeled by the vertical dot-dashed line in panel c. We only consider the portions of CLM between the two lines. The half mass of the cloud is indicated by the horizontal dotted line in panel c. Similarly, between the two vertical dotted lines, the slope is consistently 1 sigma above of the mean of the ${\rm slope}_{\rm ref}$  from all the clouds in our study(figures~\ref{fig:Figure2} -~\ref{fig:Figure16}). They are used to study whether the mass ratio is sensitive to how the CLM edges are defined, see section 2 for details. Both pairs of lines are extended across panel c and a to mark the two ends of the CLMs profile and the boundaries of the cloud length respectively. 
{\bf Panel g \& h} : A zoom of panel d towards the two ends of the slope profile. The two horizontal dot-dashed line shows respectively the 3-sigma (1.02) and 1-sigma (0.61) above the mean of all ${\rm slope}_{\rm ref}$. 
}
    \label{fig:Figure2}
\end{figure*}

\begin{figure*}
	\rotatebox[origin=c]{-90}{\includegraphics[width=16.5cm,height=22cm]{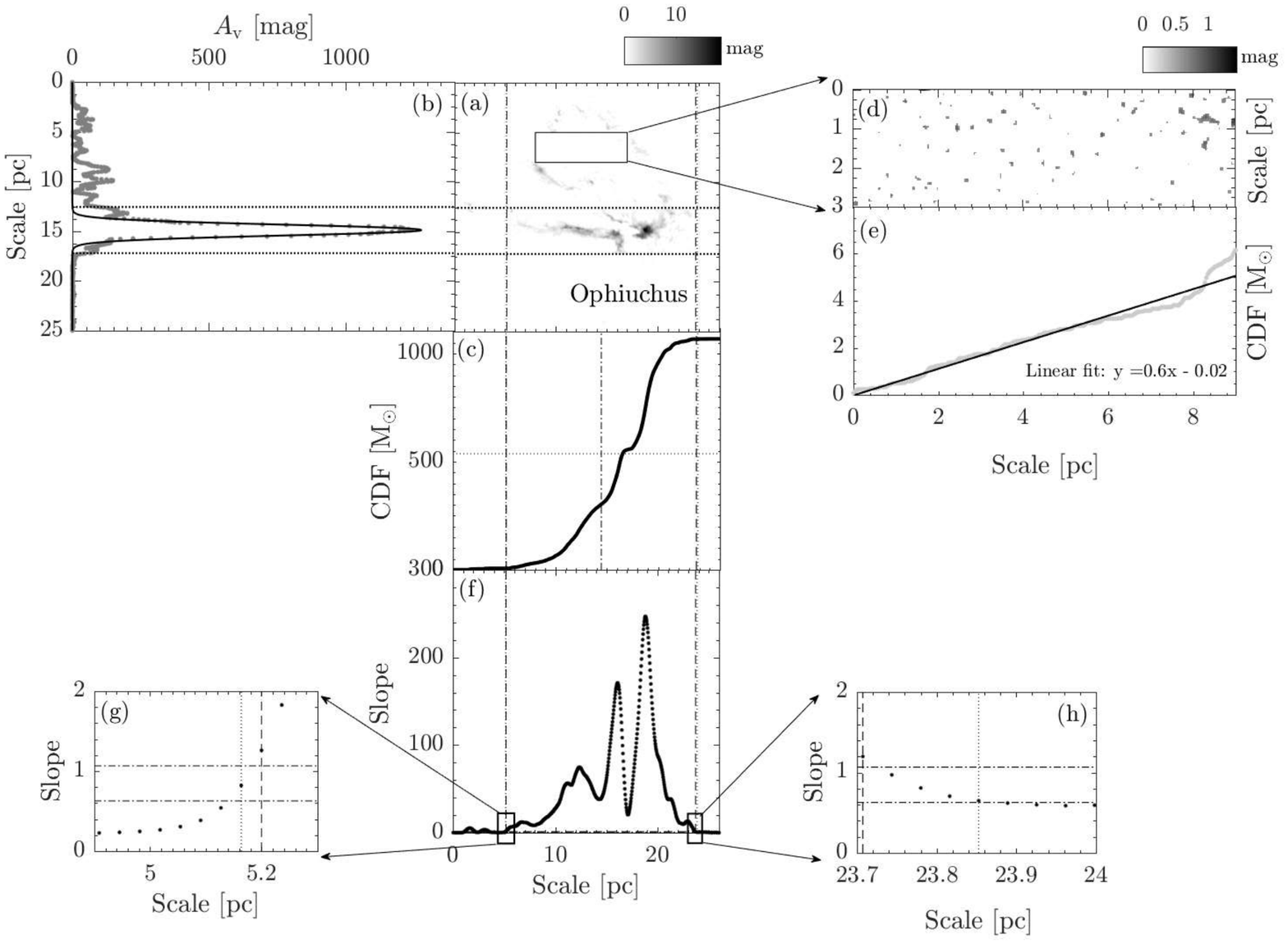}}
    \caption{Similar to figure 2, but for Ophiuchus.}
    \label{fig:Figure3}
\end{figure*}

\begin{figure*}
	\rotatebox[origin=c]{-90}{\includegraphics[width=14.5cm,height=20cm]{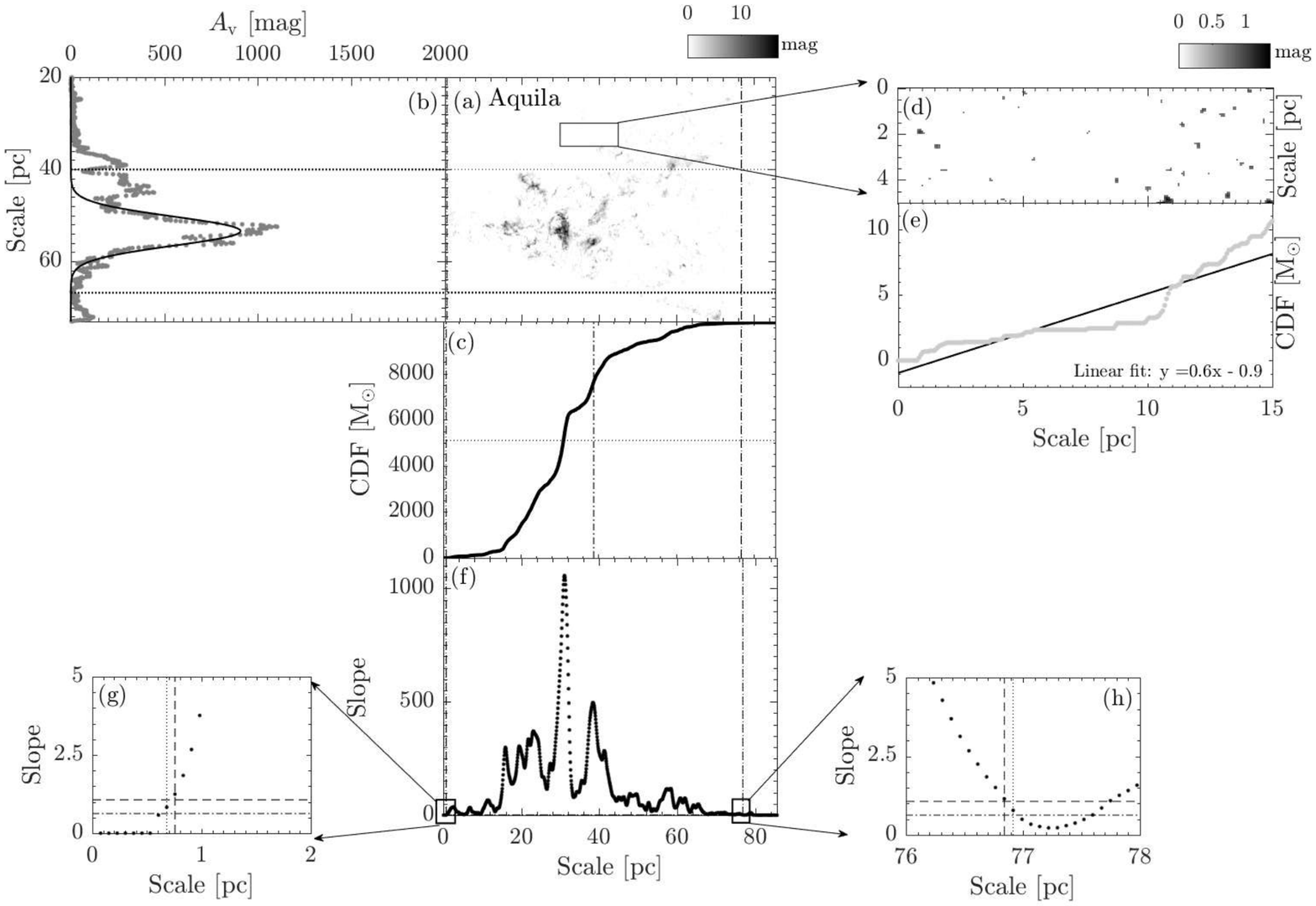}}
    \caption{Similar to figure 2, but for Aquila}
    \label{fig:Figure4}
\end{figure*}

\begin{figure*}
	\rotatebox[origin=c]{-90}{\includegraphics[width=15.5cm,height=20cm]{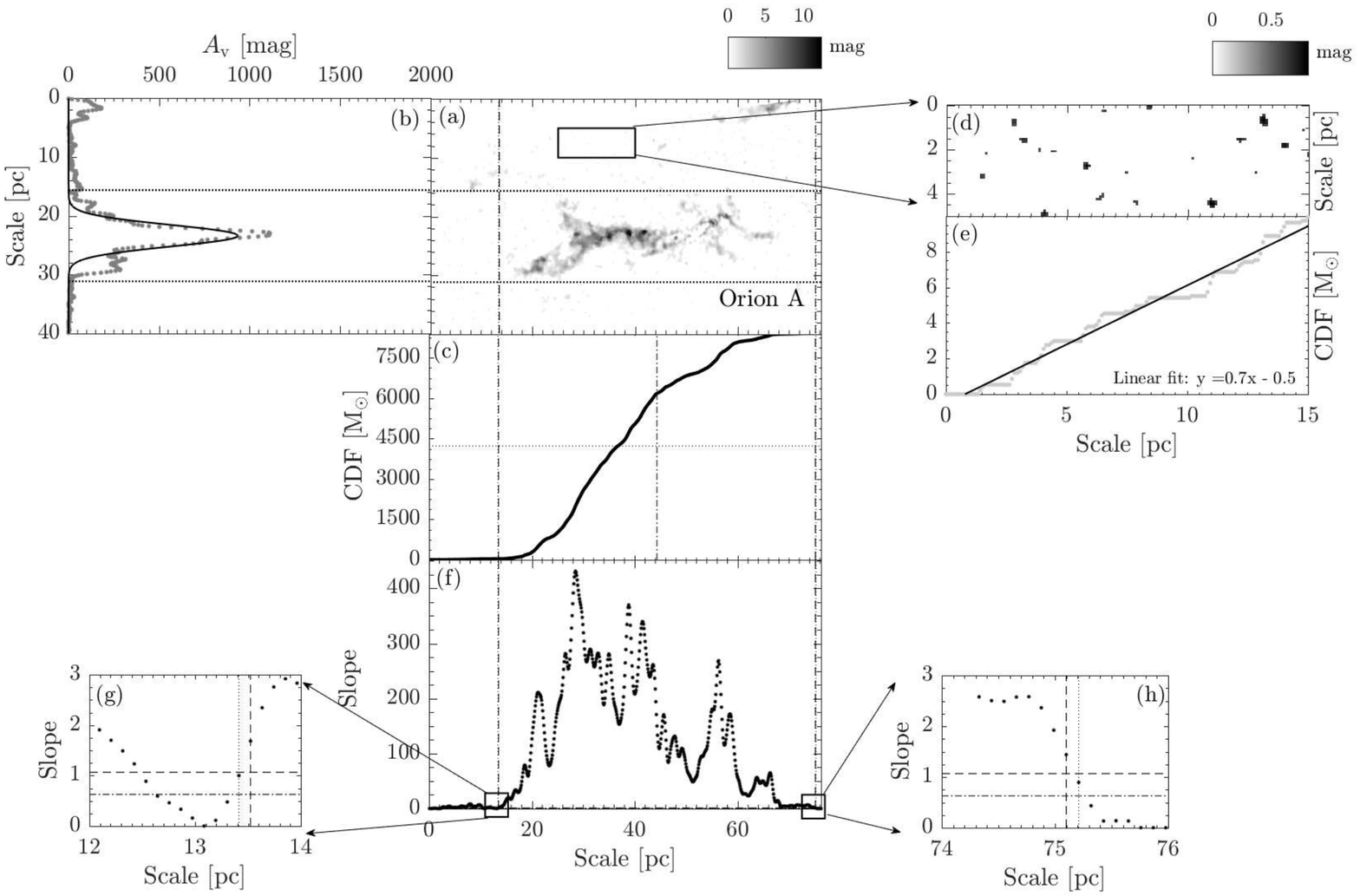}}
    \caption {Similar to figure 2, but for Orion A}
    \label{fig:Figure5}
\end{figure*}

\begin{figure*}
	\rotatebox[origin=c]{-90}{\includegraphics[width=15.5cm,height=20cm]{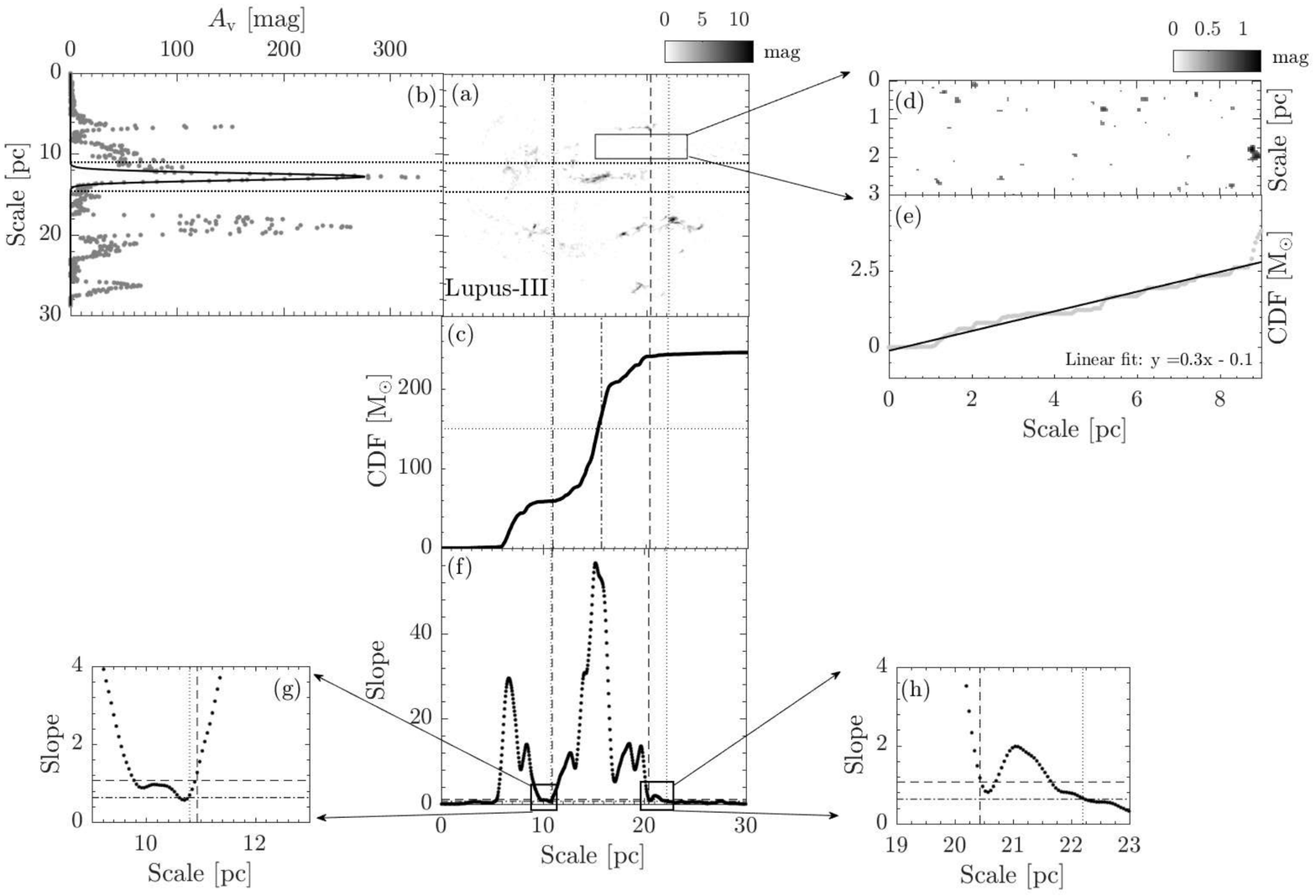}}
    \caption {Similar to figure 2, but for Lupus-III}
    \label{fig:Figure6}
\end{figure*}

\begin{figure*}
	\rotatebox[origin=c]{-90}{\includegraphics[width=15.5cm,height=20cm]{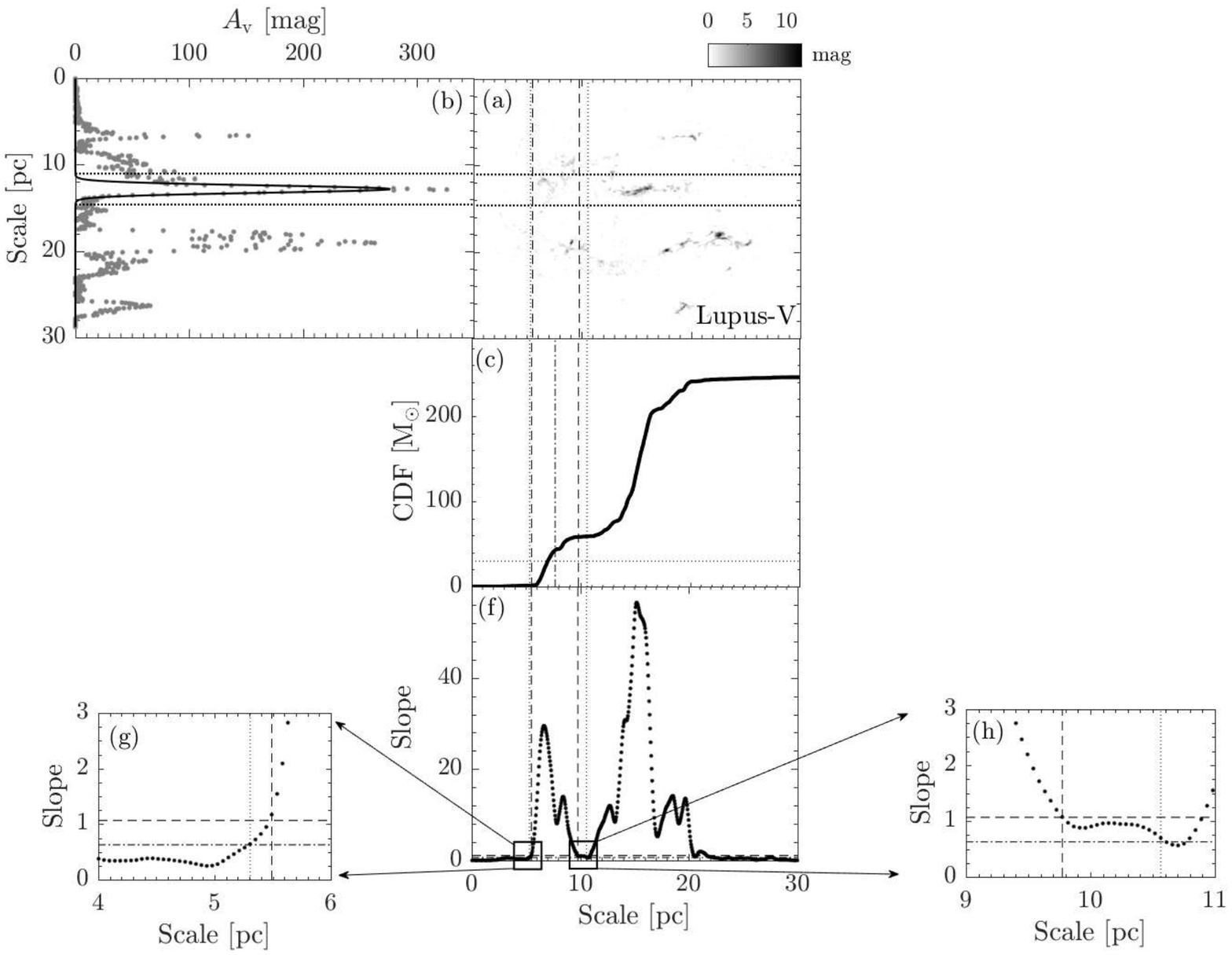}}
    \caption{Similar to figure 2, but for Lupus-V. Notice that it shares the same reference region(panel d and e) as Lupus-III. }
    \label{fig:Figure7}
\end{figure*}

\begin{figure*}
	\rotatebox[origin=c]{-90}{\includegraphics[width=15.5cm,height=20.5cm]{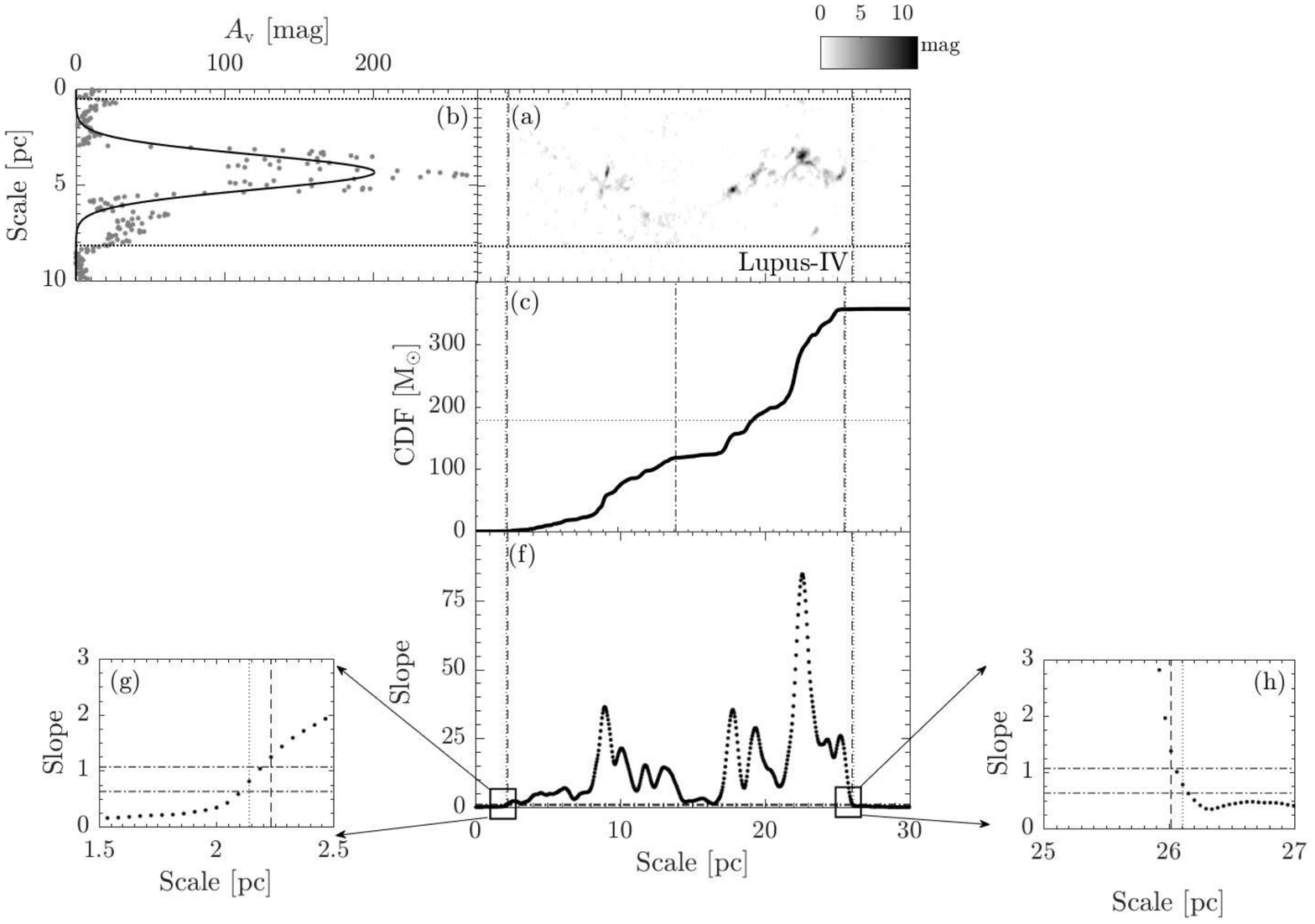}}
    \caption{Similar to figure 2, but for Lupus-IV. Notice that it shares the same reference region(panel d and e) as Lupus-III. }
    \label{fig:Figure8}
\end{figure*}

\begin{figure*}
	\rotatebox[origin=c]{-90}{\includegraphics[width=15.5cm,height=20.5cm]{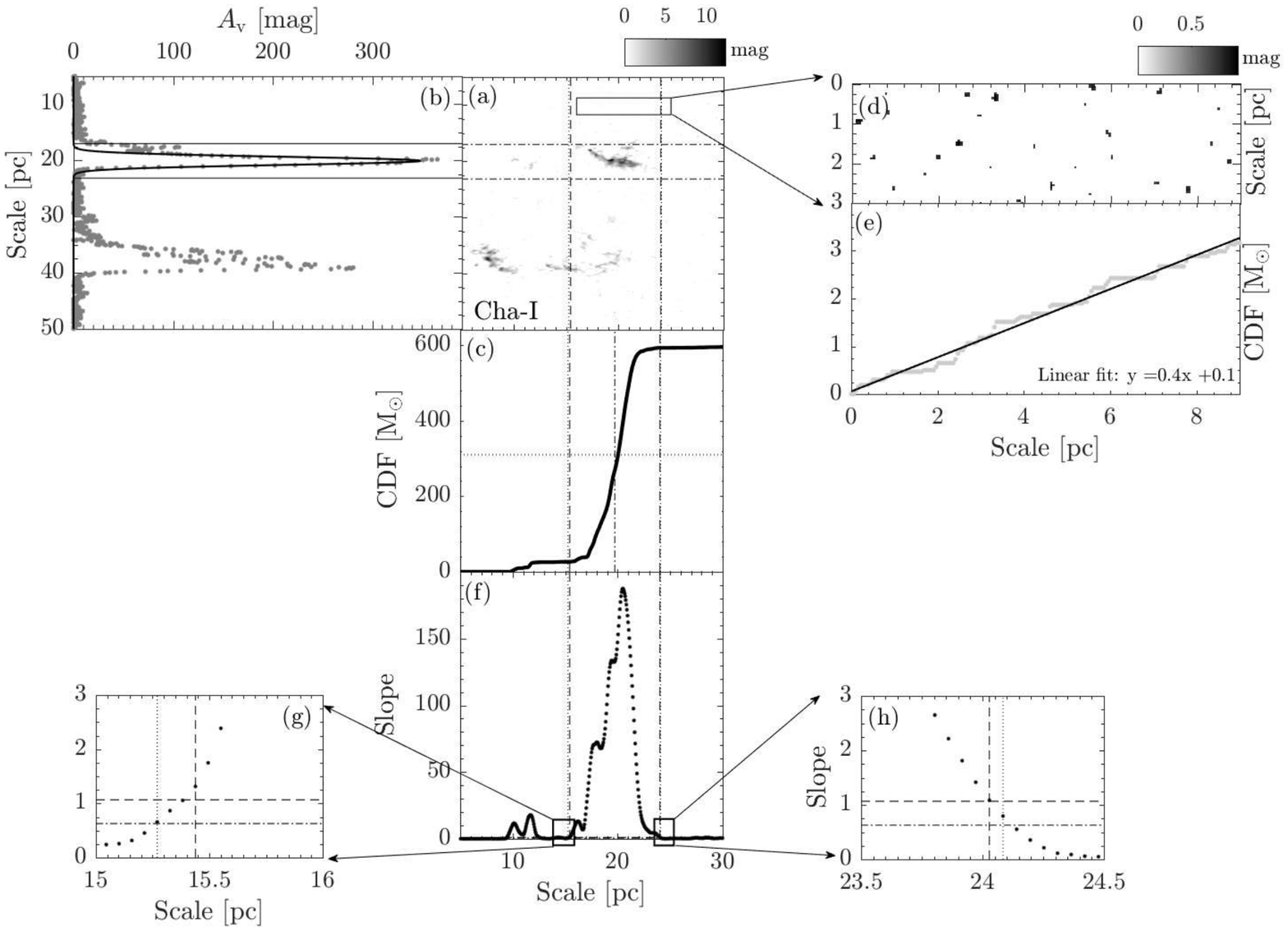}}
    \caption{Similar to figure 2, but for Cha-I.}
    \label{fig:Figure9}
\end{figure*}

\begin{figure*}
	\rotatebox[origin=c]{-90}{\includegraphics[width=15.5cm,height=20.5cm]{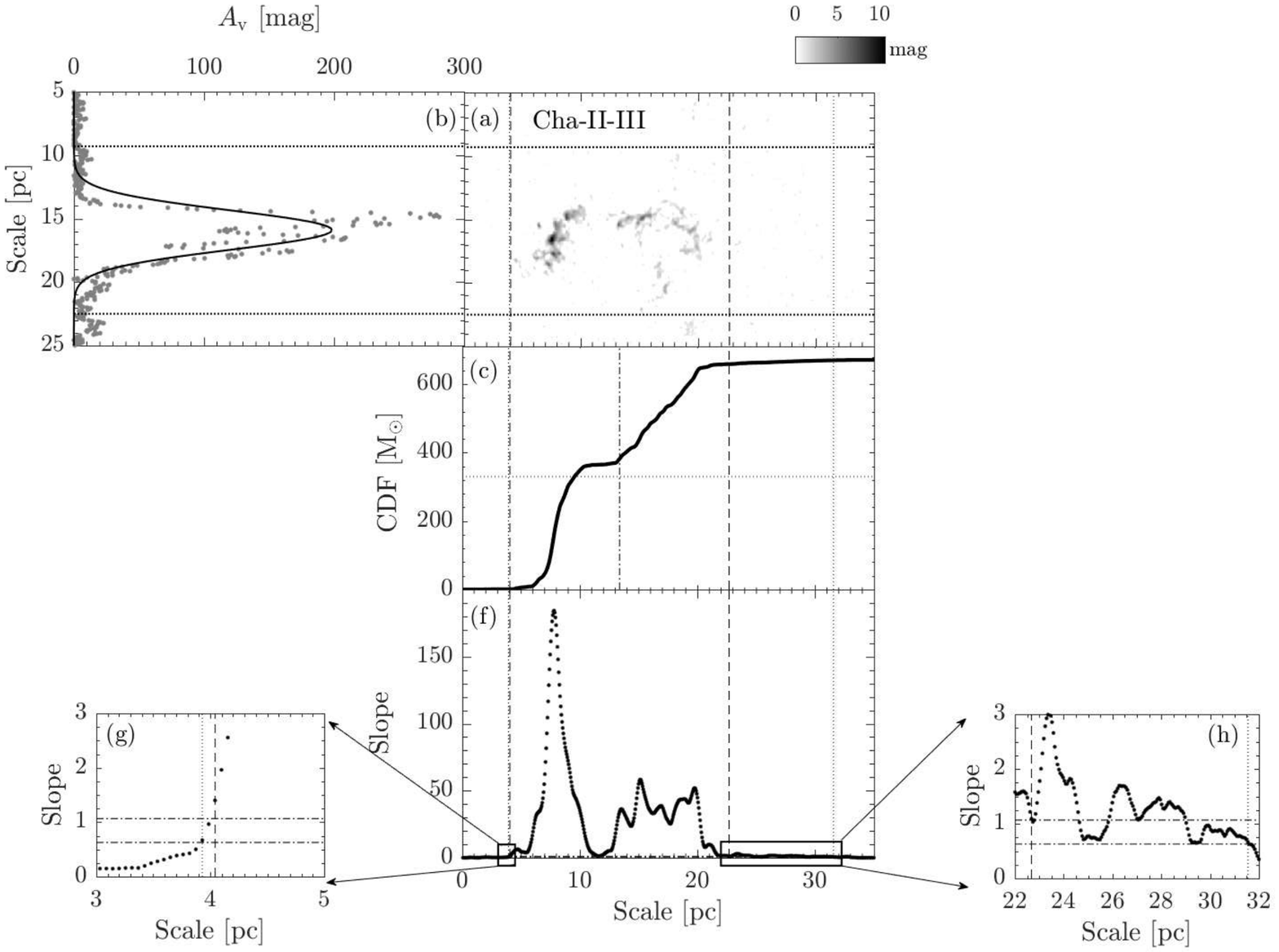}}
     \caption{Similar to figure 2, but for Cha-II-III. Notice that it shares the same reference region(panel d and e) as Cha-I. }
    \label{fig:Figure10}
\end{figure*}

\begin{enumerate}
\item{ Each cloud extinction map is rotated clockwise by the directional angle of the long axis defined in L17, such that the $x$-axis of the map aligns with the long axis of the cloud. The procedure is performed by MATLAB (ver. 2016a) built-in ``imrotate'' function (Figure~\ref{fig:Figure2} panel a). } 

\item{ To determine the cloud width, we construct the linear mass along the short axis of the extinction map by summing up pixels in the long axis direction at each y-position with extinction above the threshold defined by the N-PDF of the cloud. Panel b shows the linear mass along the short axis of CrA. In general, the linear mass profile can be approximated by a Gaussian. The half-width of the cloud is defined by 3-sigma of the Gaussian fit. As an example, the width of CrA is indicated by the horizontal solid lines.(Figure~\ref{fig:Figure2}, panel a \& b)

Some clouds show multiple peaks that locate beyond the width. Any peak has value at least half of the first structure is identified as a separated structure, and the width is defined again in the same manner as described above. All structures identified in the same extinction map are assumed to have the same long axis orientation as the overall cloud structure indicated in Table 1. }

\item{ To determine the cloud length, we construct the linear mass along the long axis: the linear mass at an $x$-position is calculated by summing over the pixels along the y-direction within the width defined in (ii) and with densities above the transition extinction. The cumulative linear mass (CLM) profile is cumulated from the left end of the map.} Figure~\ref{fig:Figure2} (panel c) presents the CLM profile of CrA.

The pixel mass is computed based on the equation below from \cite{2010ApJ...723.1019H}:
\begin{equation}
    M_{\rm cloud} = \mu\,m_{\rm H} \left(\frac{N_{\rm H}}{A_{\rm V}}\right) \times  A_{\rm V} \times A_{\rm pixel}
    \label{eq:Heiderman}
\end{equation}
where $\mu = 1.37 $ is the mean atomic mass, $m_{\rm H}$ is the mass of hydrogen, $A_{\rm pixel} = D({\rm cm})^{2}\times R(rad)^{2}$ is the area of one pixel, $D$ is the distance of the cloud in centimeters, $R$ is the pixel size in radian, and $N_{\rm H}/A_{\rm V} = 1.37 \times 10^{21}\;{\rm cm^{-2}\;mag^{-1}}$ is the conversion factor between visual extinctions and column density. The error of mass of a pixel is based on the uncertainty in the extinction of the corresponding pixel from \cite{2011PASJ...63S...1D}.

\item{The two ends of a long-axis linear mass are decided by comparing its CLM with the CLM from regions beyond the cloud width(the rectangular box in panel a), which we call ${\rm CLM}_{\rm ref}$.  We require the CLM slope within the two ends to be consistently above a threshold which is computed based on the slopes of ${\rm CLM}_{\rm ref}$ (${\rm slope}_{\rm ref}$). The slopes are defined in the follows. For each position along the CLM profile, we define the slope by a linear fit of the CLM profile over a one-parsec interval centered at the position. Figure~\ref{fig:Figure2} (panel d) illustrates the slope profile of CrA. The reference region and the corresponding ${\rm CLM}_{\rm ref}$ of CrA is shown in Figure~\ref{fig:Figure2} (panel d and e). For molecular clouds within 250pc, each reference region has a dimension of $3 {\rm pc} \times 9 {\rm pc}$. For molecular clouds further than 250 pc, as the physical resolution changes, we enlarge the size of each region to $5{\rm pc} \times 15 {\rm pc}$ in order to include sufficient pixel numbers. ${\rm CLM}_{\rm ref}$ of each reference region is calculated and linearly fitted for the slope. We take the fitted slope as ${\rm slope}_{\rm ref}$. 3-sigma above the mean of the ${\rm slope}_{\rm ref}$ from all clouds is used to define the ends of each cloud. The two ends of CrA are indicated by the solid vertical dashed lines.(Figure~\ref{fig:Figure2} panel a,c \& d). Any peak beyond the ends with value at least half of the first structure is identified as a separated structure and the length is defined again based on the aforementioned criteria.}

\item{The resultant map will be used to study the evenness of mass distribution. The middle point of the cloud is halfway between the two ends of the CLM. We quantify the evenness of mass distributions by the mass ratio between the two halves of the linear mass.}

\end{enumerate}
The resulting maps following procedures (i) to (v) are shown in Figure ~\ref{fig:Figure2} -~\ref{fig:Figure16}.

\begin{table}
\centering
\caption{Gould Belt Clouds distances, long-axis orientations and magnetic field orientations}
\label{my-label}
\begin{tabular}{|c|c|c|c|}
\hline
Cloud name       & \begin{tabular}[c]{@{}c@{}}Distance\\ (pc)\end{tabular} & \begin{tabular}[c]{@{}c@{}}Long-axis\\ (degree)\end{tabular} & \begin{tabular}[c]{@{}c@{}}B-field$^{*}$\\ (degree)\end{tabular} \\ \hline
Taurus           & 135$\pm$20 $^{a}$                                                     & 75                                                           & 1$\pm$18                                                  \\ \hline
Pipe Nebula      & 130$\pm$18 $^{b}$                                                     & -45                                                          & 59$\pm$13                                                 \\ \hline
Chameleon        & 193$\pm$39 $^{c\#}$                                                     & 22                                                           & -76$\pm$11                                                \\ \hline
Musca       & 160$\pm$25 $^{c}$                                                     & 27                                                           & -82$\pm$8                                                \\ \hline
Lupus-I          & 150$\pm$20 $^{c}$                                                    & -1                                                           & 86$\pm$13                                                 \\ \hline
Orion-B         & 398$\pm$12 $^{b}$                                                    & -30                                                           & 93$\pm$20                                                 \\ \hline
IC5146         & 460$\pm$80 $^{d}$                                                    & -38                                                           & 67$\pm$16                                                 \\ \hline
Ophiuchus        & 125$\pm$18 $^{a}$                                                    & -45                                                          & -81$\pm$25                                                \\ \hline
Corona Australis & 130$\pm$25 $^{c}$                                                    & -26                                                          & -26$\pm$32                                                \\ \hline
Lupus II-VI      & 160$\pm$40 $^{c\#}$                                                     & -73                                                          & 82$\pm$11                                                 \\ \hline
Aquila           & 260$\pm$10 $^{c}$                                                    & -75                                                          & -67$\pm$10                                                \\ \hline
Orion-A         & 378$\pm$10 $^{b}$                                                    & 83                                                           & 59$\pm$25                                                 \\ \hline
\end{tabular}
\begin{tablenotes}
      \small
      \item  $^{a}$ \cite{2014ApJ...786...29S}.$^{b}$ \cite{2010ApJ...724..687L}.$^{c}$\cite{2010ApJ...723.1019H}$^{d}$ \cite{2011A&A...529L...6A}$^{\#}$ The distance of cloud is computed by the averages between the distances to Cha I, II \& III, and Lupus II, III, IV, V \& VI respectively. $^{*}$ Mean B-field directions based on PLANCK data are adopted from L17. The magnetic field directions of the first seven clouds have axis-field offsets greater than $70$ degrees, while the latter five have the offsets less than $30$ degrees. 
    \end{tablenotes}
\end{table} 

\section{Results and Discussion}

For a molecular cloud with even mass distributions, the cloud CLM should reach $50\%$ (dotted line in Figure~\ref{fig:Figure2} panel c) at the half-length (dot-dashed line in Figure~\ref{fig:Figure2} panel c) of the filament. The more uneven is the mass distribution, the further the CLM should deviate from the point of the intersection between the dot-dashed and the dotted lines. 

\subsection*{Statistical analysis}

Table 2 summarizes the mass ratio for the 15 regions. Figure 17 presents the mass ratio versus cloud-field direction offset. We also test whether the results above depend on how the CLM edges are defined by defining the ends of all the CLMs using a lower threshold, which is 1-sigma above the mean  of all ${\rm slope}_{\rm ref}$. The corresponding cloud ends are denoted by the vertical dotted lines in Figure~\ref{fig:Figure2} (panel a,c \& d); the result is also shown in Figure17. Figure~\ref{fig:Figure17} seems to show a positive correlation, that molecular clouds closer to a perpendicular alignment tend to have more even mass distributions. The trend is independence from the latitude of the cloud or the choices of the threshold. In the following, we present the permutation test and Spearman rank correlation (SRC) test to investigate the significance of ``bimodality'' and positive correlation between mass ratio and cloud-field direction offsets.

\begin{table}
\caption{The cloud-field direction offsets and mass ratio of the 15 molecular clouds.}
\begin{tabular}{|c|c|c|}
\hline
Cloud & \begin{tabular}[c]{@{}c@{}}cloud-field direction offsets$^{a}$\\ (deg)\end{tabular} & Mass ratio$^{b}$ \\ \hline
Corona & 0  & $0.19\,(0.19)$ \\ \hline
Ophiuchus & 36 & $0.39\,(0.39)$ \\ \hline
Aquila & 8 & $0.34\,(0.34)$ \\ \hline
Orion A & 24 & $0.34\,(0.34)$ \\ \hline
Lupus III & 25  & $0.72\,(0.27)$ \\ \hline
Lupus IV & 25  & $0.49\,(0.49)$ \\ \hline
Lupus V & 25 & $0.39\,(0.35)$ \\ \hline
Cha I & 82 & $0.75\,(0.72)$ \\ \hline
Cha II-III & 82 & $0.73\,(0.24)$ \\ \hline
Pipe & 76  & $0.69\,(0.69)$ \\ \hline
Musca & 71 & $0.85\,(0.65)$ \\ \hline
IC5146 & 75 & $0.70\,(0.70)$ \\ \hline
Lupus I & 85  & $0.55\,(0.55)$ \\ \hline
Orion B & 57 & $0.81\,(0.75)$ \\ \hline
Taurus & 74 & $0.76\,(0.83)$ \\ \hline

\end{tabular}
\begin{tablenotes}
      \small
      \item  $^{a}$The cloud-field direction offset is computed by taking the differences between the long axis orientations and B-field directions in table 1. $^{b}$ The mass ratios depend on the cloud threshold. Two thresholds , "3-sigma" and "1-sigma", are used as described in section 2. The results from 1-sigma are shown in the parentheses.
 
    \end{tablenotes}
\end{table}

For the permutation test, the molecular clouds are divided into two groups, with the sample sizes equal to the numbers of clouds with parallel (${\rm N}_{\parallel} = 7$) and perpendicular (${\rm N}_{\perp} = 8$) alignment. The corresponding population means of the mass ratio are $\mu_{\parallel}$ and $\mu_{\perp}$ respectively. We adopt a null hypothesis $H_{0} : \mu_{\perp} = \mu_{\parallel}$, and the alternative hypothesis $H_{1} : \mu_{\parallel} < \mu_{\perp}$. To test the hypotheses, we reassign the clouds into two groups ($G_{1}$,$G_{2}$) with the sample size of 7 and 8 respectively, and compute the difference in the means of the two groups($D = \mu_{G_{1}} -\;\mu_{G_{2}}$).  This step is repeated for all possible re-groupings, which will obtain $^{15}C_{7}$ (``15 choose 7'') values of D's. We use the built in MATLAB ``combnk'' function to enumerate all combinations. Each combination is compared with the difference between the observed means ($D_{\rm obs} = \mu_{\parallel}-\mu_{\perp}$). The percentage of occurrences of $D \leq D_{\rm obs}$ provide an estimate of the $p$-value. The null hypothesis is rejected if the $p$-value is smaller or equal to $0.05$. Based on the permutation test, the corresponding $p$-values is $0.0006$, and $0.0017$ for the case with the lower threshold. The results support a bimodal distribution of mass ratio.

We study the significance of a positive correlation by Spearman rank correlation (SRC) test. MATLAB ``$mult-comp-perm-corr$'' function is used to perform the SRC test. The SRC coefficient reflects how well two variables correlate based on their ranks instead of the actual numerical values \citep{WalpoleMyersMyersEtAl2011}. Such a nonparametric test has no assumptions on the underlying distributions and is insensitive to outliers. For this work, the SRC is between the rank of the mass ratio ($rank_{\rm ratio}$), and the rank of cloud-field direction offset ($rank_{\rm angle}$). The SRC ranges between $\pm1$. The more positive/negative it is, the more positively/negatively correlated the ranks are. $\rm{SRC} \sim 0$ means there is no correlation. 

For the SRC test, we adopt a null hypothesis $H_{0} : {\rm SRC}(rank_{\rm ratio}, rank_{\rm angle}) = 0$, and the alternative hypothesis $H_{1} : {\rm SRC}(rank_{\rm ratio}, rank_{\rm angle}) > 0$. The $rank_{\rm ratio}$ is randomly permuted to generate $rank_{{\rm ratio,perm}}$. We perform random permutation for $10^{6}$ times. The $p$-value is thus estimated by the Plum frequency of (${\rm SRC}(rank_{{\rm ratio,rand}}$ , $rank_{\rm angle}) > {\rm SRC}_{\rm obs}$). The corresponding $\rm {SRC}_{obs}$($p$-value) is $0.63$(0.006), and 0.54($0.019$) for the case with a lower threshold. Both results suggest the observed trends in Figure~\ref{fig:Figure17} is also consistent with a positive correlation.

\section{Summary and Outlooks}
We present an analysis of the near-infrared dust extinction maps of twelve Gould Belt molecular clouds. Inspired by the findings in paper-I and L17, we study whether the cloud-field direction offset has any effect on the cloud linear mass distribution. The perpendicular clouds show more even mass distribution. This is consistent with the conclusion from L17 that magnetic fields perpendicular to clouds can hinder the cloud fragmentation more efficiently and thus lead to a lower star formation efficiency.

The technique utilised in this work to identify the transition threshold from the molecular cloud N-PDFs may provide a simple yet relatively robust approach compared to the conventional methods that involve log-normal fitting, for which the choices on the fitting range usually are not well constrained. In addition to the different ways to analyze the \twomass data, depending on the selected fitting algorithm, the resulting transition density of N-PDF can vary in the order of a few $A_{\rm V}$ magnitudes (paper-I Figure 7). Our next step is to further split the regions studied here by investigating individual sub-regions above the N-PDF transition threshold and adopting the corresponding magnetic field directions from the PLANCK polarimetry data. This way we can test whether the trend observed here remains at smaller scales.

\section*{Acknowledgements}
This research is supported by four grants from the Research Grants Council of the Hong Kong:  Early Career Scheme 24300314; General Research Fund 14305717,14600915, and 14304616.




\bibliographystyle{mnras}
\bibliography{References} 

\renewcommand{\thefigure}{\arabic{figure}}
\begin{figure*}
	\rotatebox[origin=c]{-90}{\includegraphics[width=14.5cm,height=20cm]{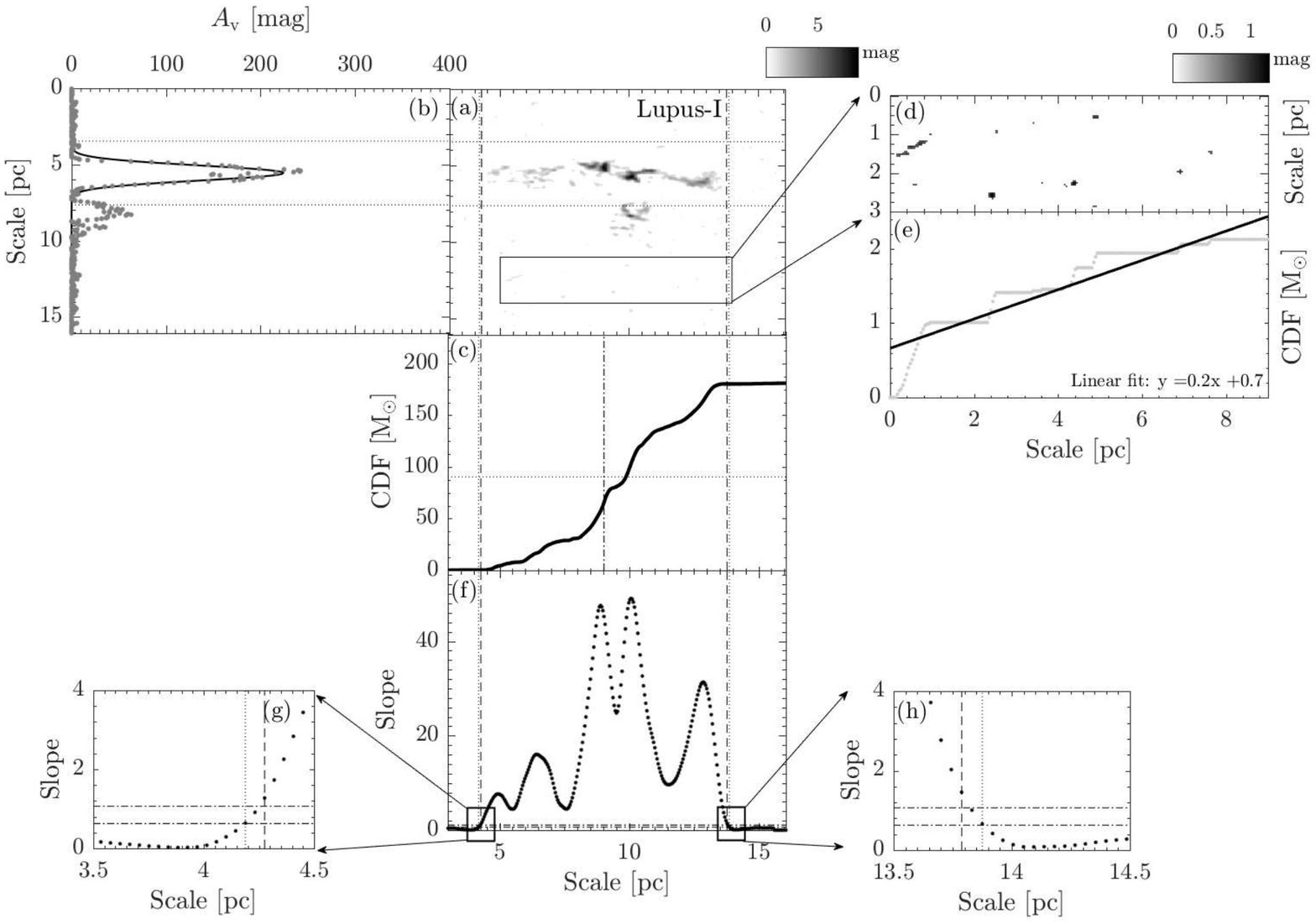}}
    \caption{Similar to figure 2, but for Lupus-I. }
    \label{fig:Figure11}
\end{figure*}
\begin{figure*}
	\rotatebox[origin=c]{-90}{\includegraphics[width=16cm,height=22cm]{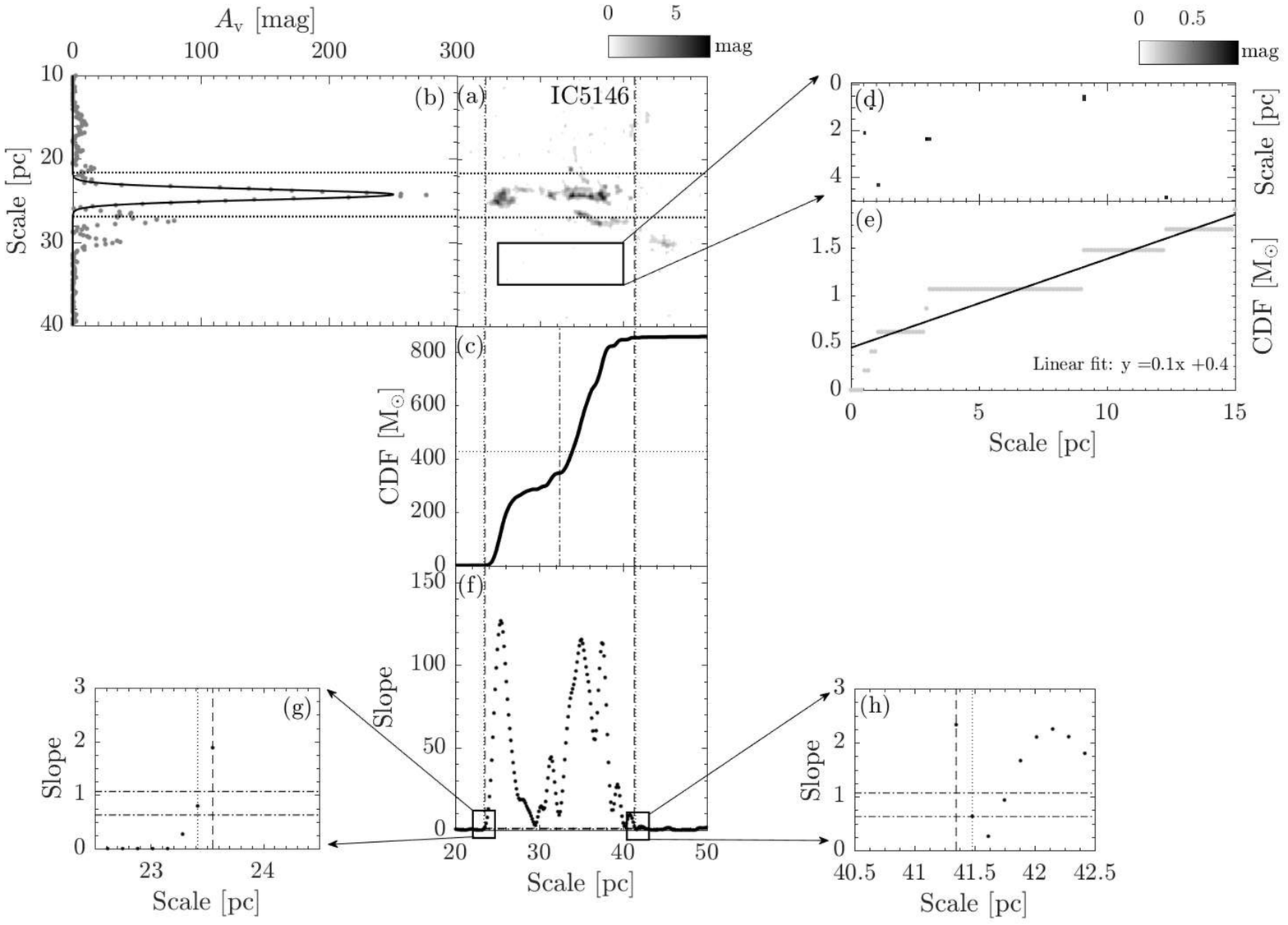}}
    \caption{Similar to figure 2, but for IC5146}
    \label{fig:Figure12}
\end{figure*}
\begin{figure*}
	\rotatebox[origin=c]{-90}{\includegraphics[width=16cm,height=21cm]{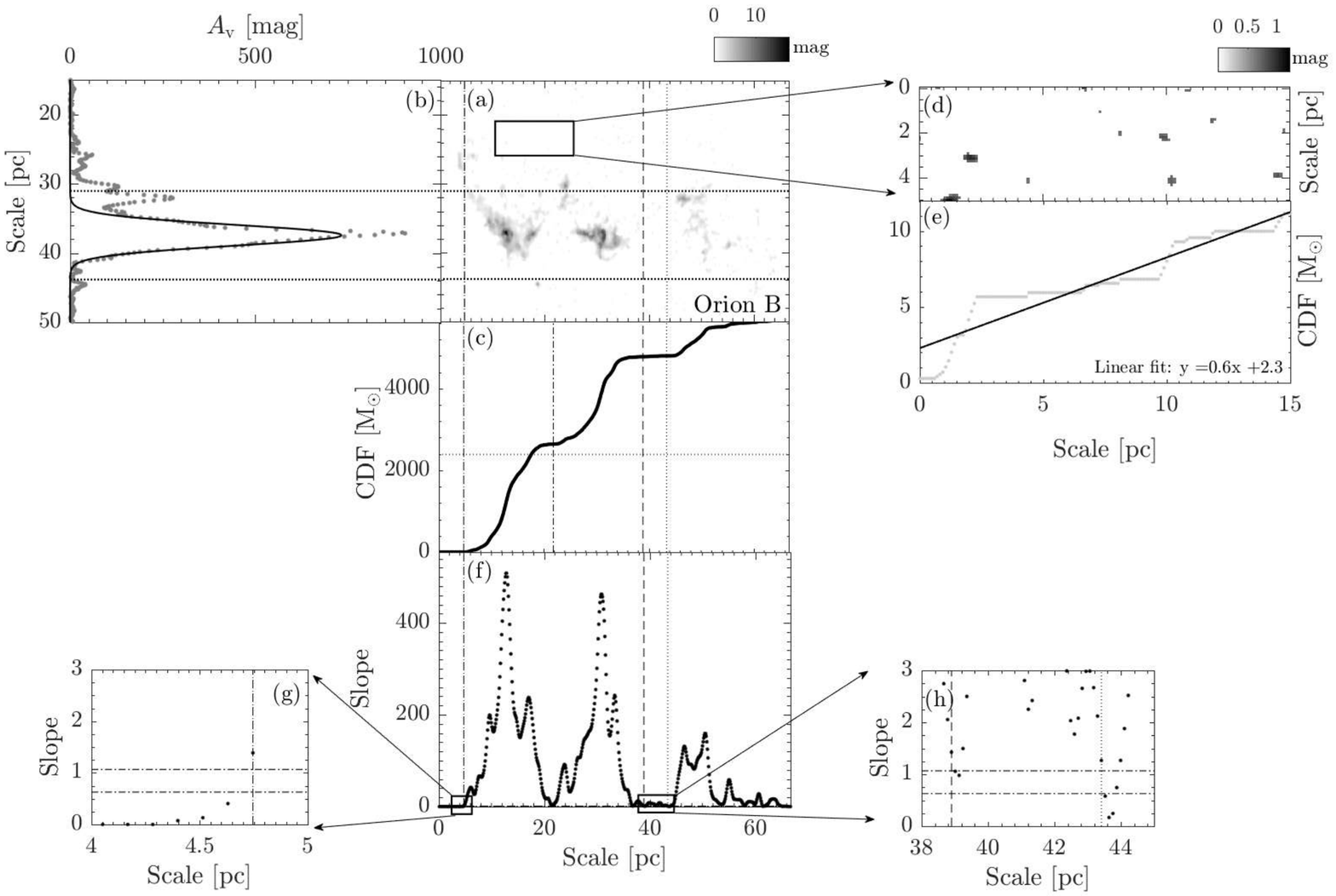}}
    \caption{Similar to figure 2, but for Orion B. }
    \label{fig:Figure13}
\end{figure*}
\begin{figure*}
	\rotatebox[origin=c]{-90}{\includegraphics[width=14.5cm,height=20cm]{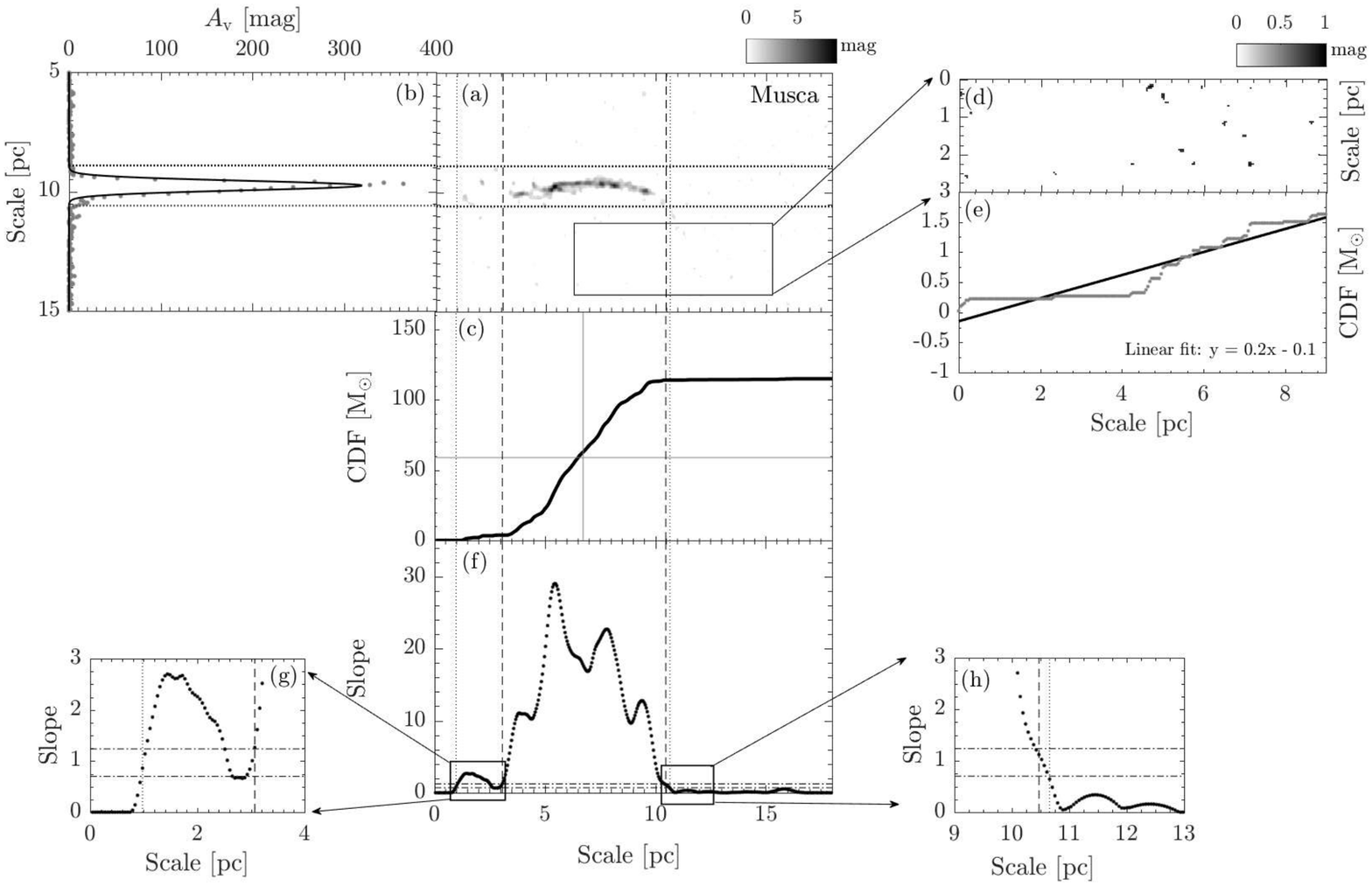}}
    \caption{Similar to figure 2, but for Musca. }
    \label{fig:Figure14}
\end{figure*}
\begin{figure*}
	\rotatebox[origin=c]{-90}{\includegraphics[width=15.5cm,height=22cm]{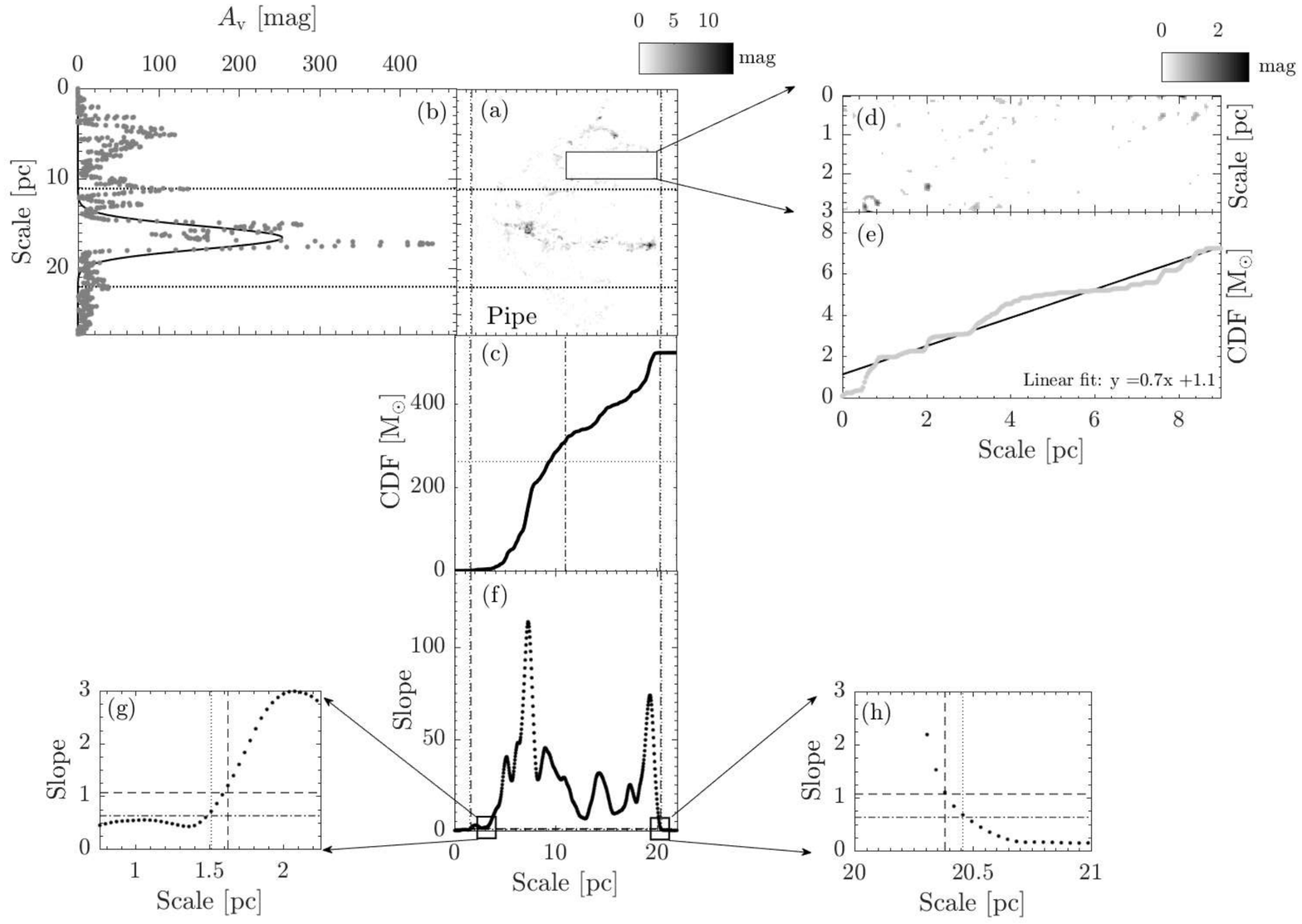}}
    \caption{Similar to figure 2, but for Pipe Nebula.}
    \label{fig:Figure15}
\end{figure*}
\begin{figure*}
	\rotatebox[origin=c]{-90}{\includegraphics[width=18cm,height=21cm]{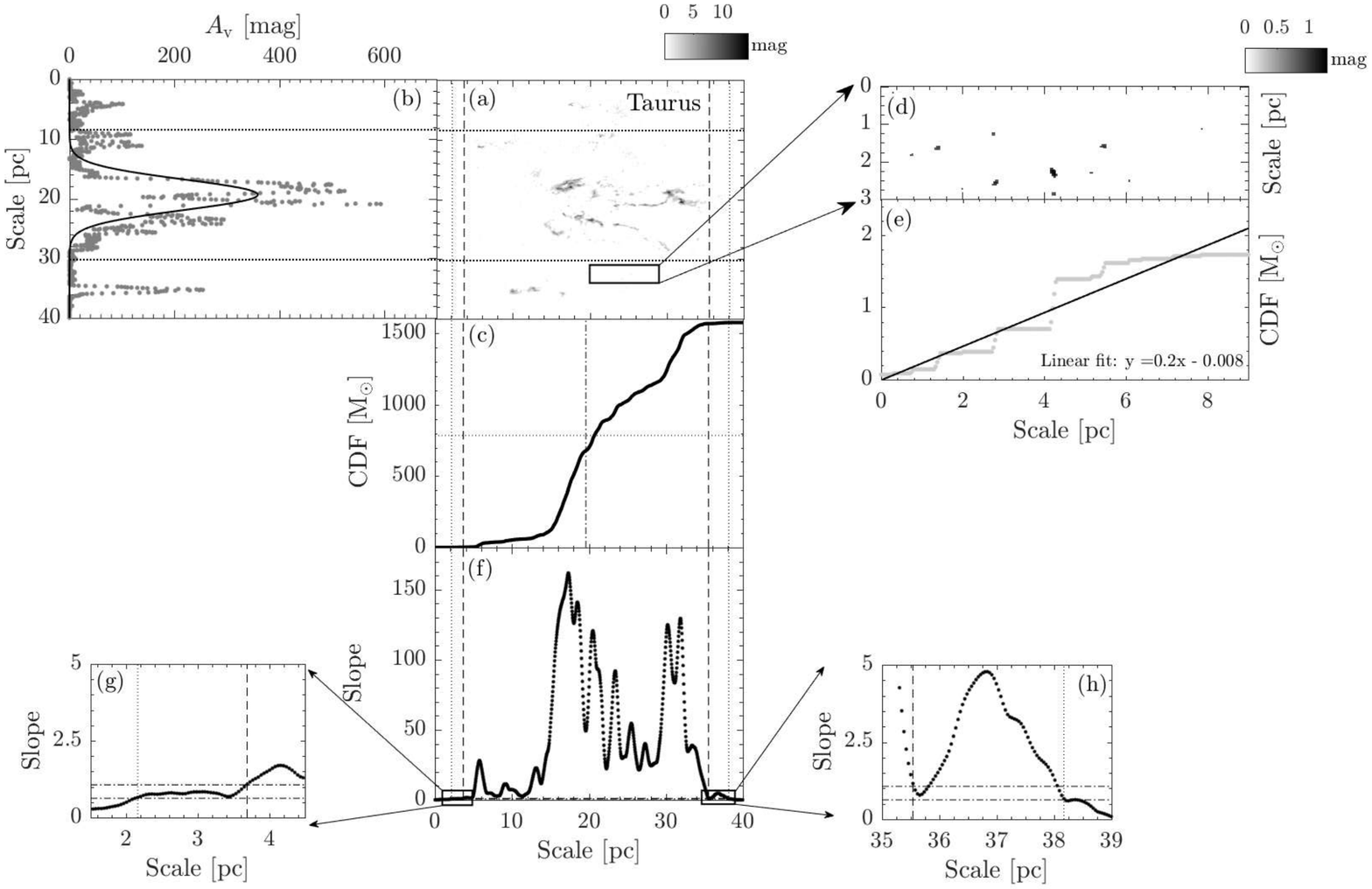}}
    \caption{Similar to figure 2, but for Taurus.}
    \label{fig:Figure16}
\end{figure*}
\begin{figure*}
	\includegraphics[width=19cm]{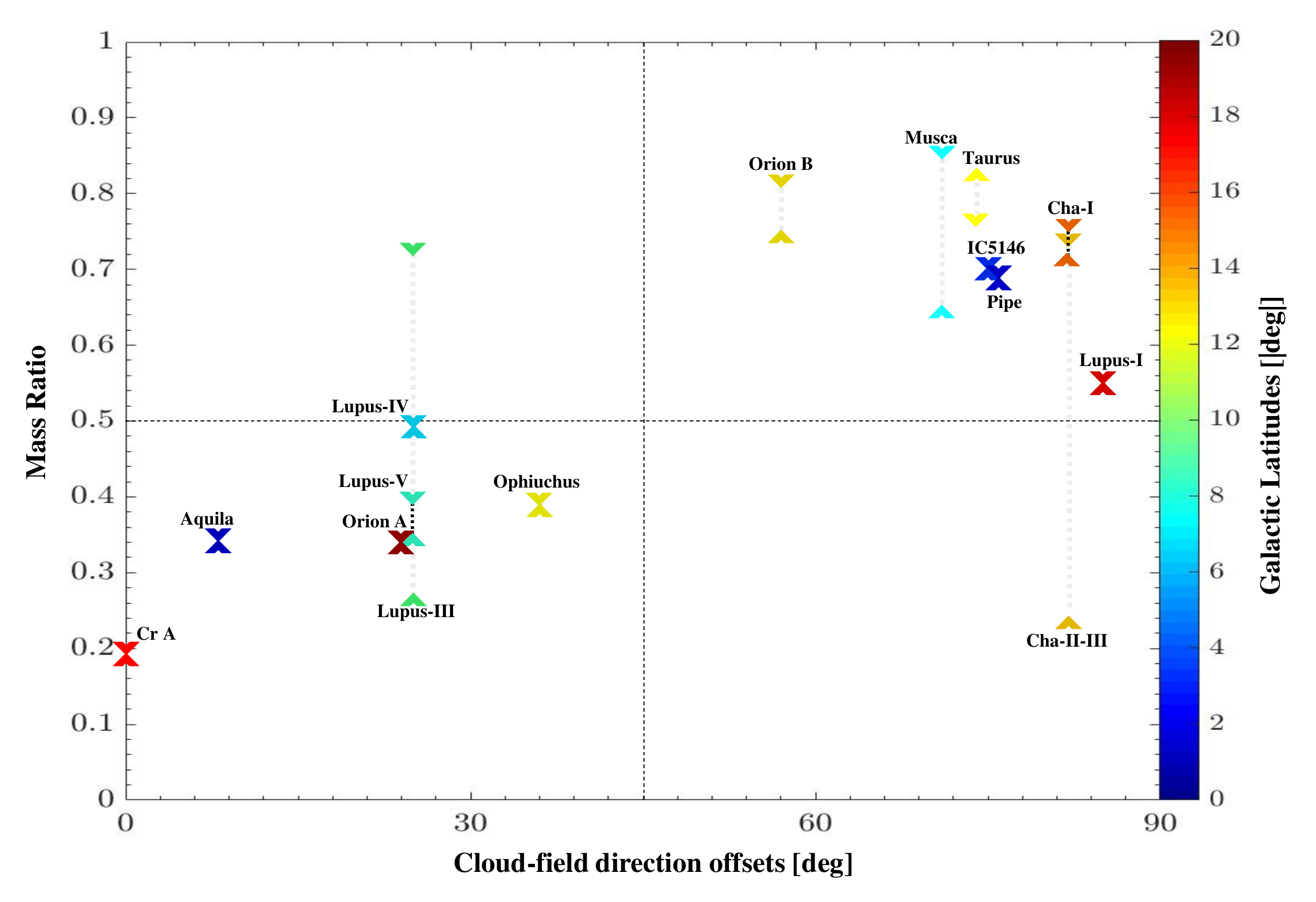}
    \caption{ Mass ratio vs. cloud-field direction offset. The ratio is between the masses from the two halves of a cloud, whose long-axis direction is measured in Paper-I, and we define a cloud by a lower threshold of the slope of the cumulative linear mass along this direction (Fig. 2-16; panel f). Two lower thresholds are adopted to demonstrate that the correlation showing here is not sensitive to the threshold. The thresholds are respectively 1-sigma (shown as "$\wedge$") and 3-sigma (shown as "v") above the mean slope in the background (see section 2 for more details). The one- and three-sigma results from one cloud are connected by a dotted line, if the respective "$\wedge$" and "v" symbols are not overlapped. Most clouds with offsets smaller/larger than 45 degree possess mass ratios below/above 0.5. The mass ratios are not sensitive to the choices of the threshold, except Lupus-III and Cha-II-III. The color of the symbol represents the lower bound latitude of the cloud. The mass ratios are independent of the latitude of the cloud.
 }
    \label{fig:Figure17}
\end{figure*}



\clearpage
\appendix
\clearpage

\section{Transition extinction based on N-PDF of molecular cloud}
In this section, we describe the method that we use to locate the transition density of N-PDFs. We identify the transition density by the slope of the binned N-PDFs. At extinction value below the peak, the slope is relatively constant. The slope of the N-PDF decreases near the peak extinction, then further decreases to negative as the extinction increases. However, when there are regions with gravitation contraction, the N-PDF deviates from the log-normal distribution to power-law at the transition extinction \citep[also see e.g.][for other alternative explanations]{2017A&A...606L...2A}, where the slope of the N-PDF will show a flattening or increase in slope from negative toward zero (the solid red line). In light of the change in the slope, we plot the slope distribution of the N-PDF. The transition density is chosen as the centre of the first extinction bin at which the slope starts to increase from negative towards zero. We set the bin width of the N-PDF to 0.2 mag, which is the average noise level of the extinction map according to \citet{2011PASJ...63S...1D}, and thus the lowest sensible value of the bin
width. The transition column density is defined by the bin at which
the slope is at the first minimum. Figures A1 and A2 show the N-PDFs, and the corresponding slope distributions of all Gould Belt molecular clouds analysed in this work. Table A1 summaries the transition density of all the molecular clouds in this work. The map regions are the same as in L17. The corresponding transition extinction value is defined as the extinction at which the slope first deviates from the increasing trend as described above. The transition densities span a rather tight range from 0.5 to 0.8.

\begin{figure}
	\includegraphics[width=\columnwidth,height = 17cm]{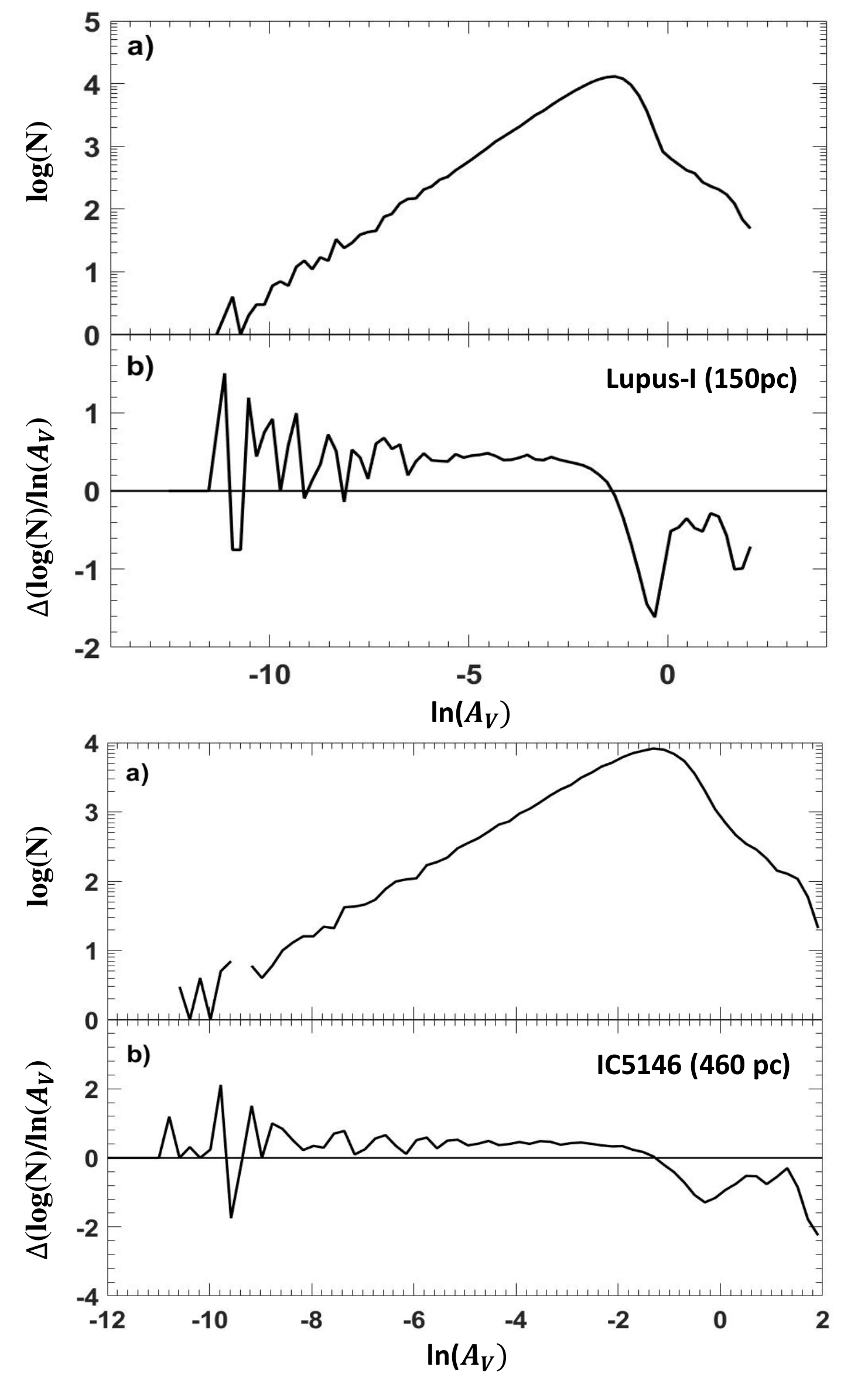}
    \caption{N-PDFs and the corresponding slope distribution of molecular clouds regions as closer to be perpendicular with B-field by L17.}
    \label{fig:FigureA1}
\end{figure}

\renewcommand{\thefigure}{\arabic{figure} (Cont.)}
\addtocounter{figure}{-1}
\begin{figure}
	\includegraphics[width=\columnwidth,height = 17cm]{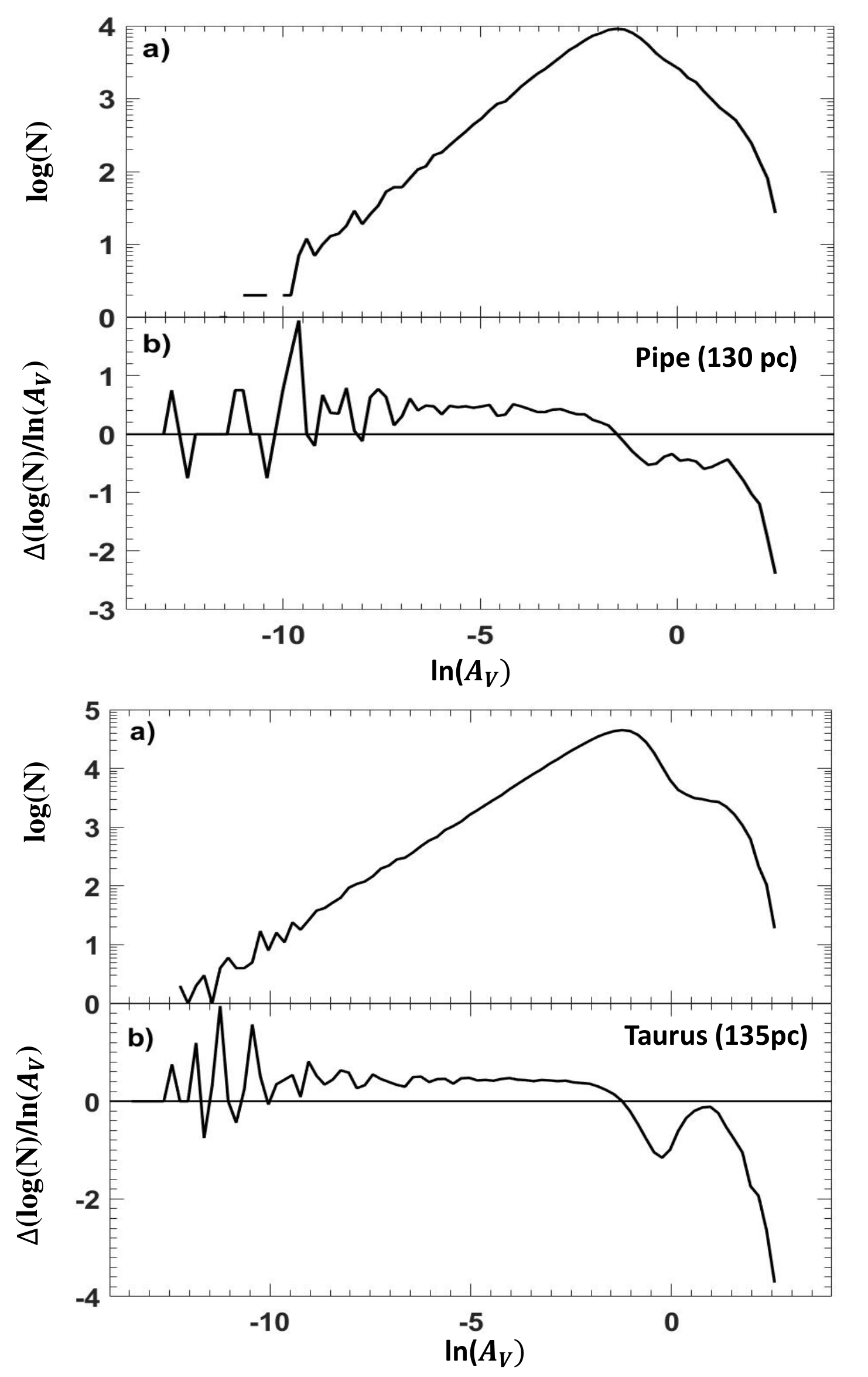}
    \caption{}
    \label{fig:FigureA2}
\end{figure}
\renewcommand{\thefigure}{\arabic{figure}}
\renewcommand{\thefigure}{\arabic{figure} (Cont.)}
\addtocounter{figure}{-1}
\begin{figure}
	\includegraphics[width=\columnwidth,height = 17cm]{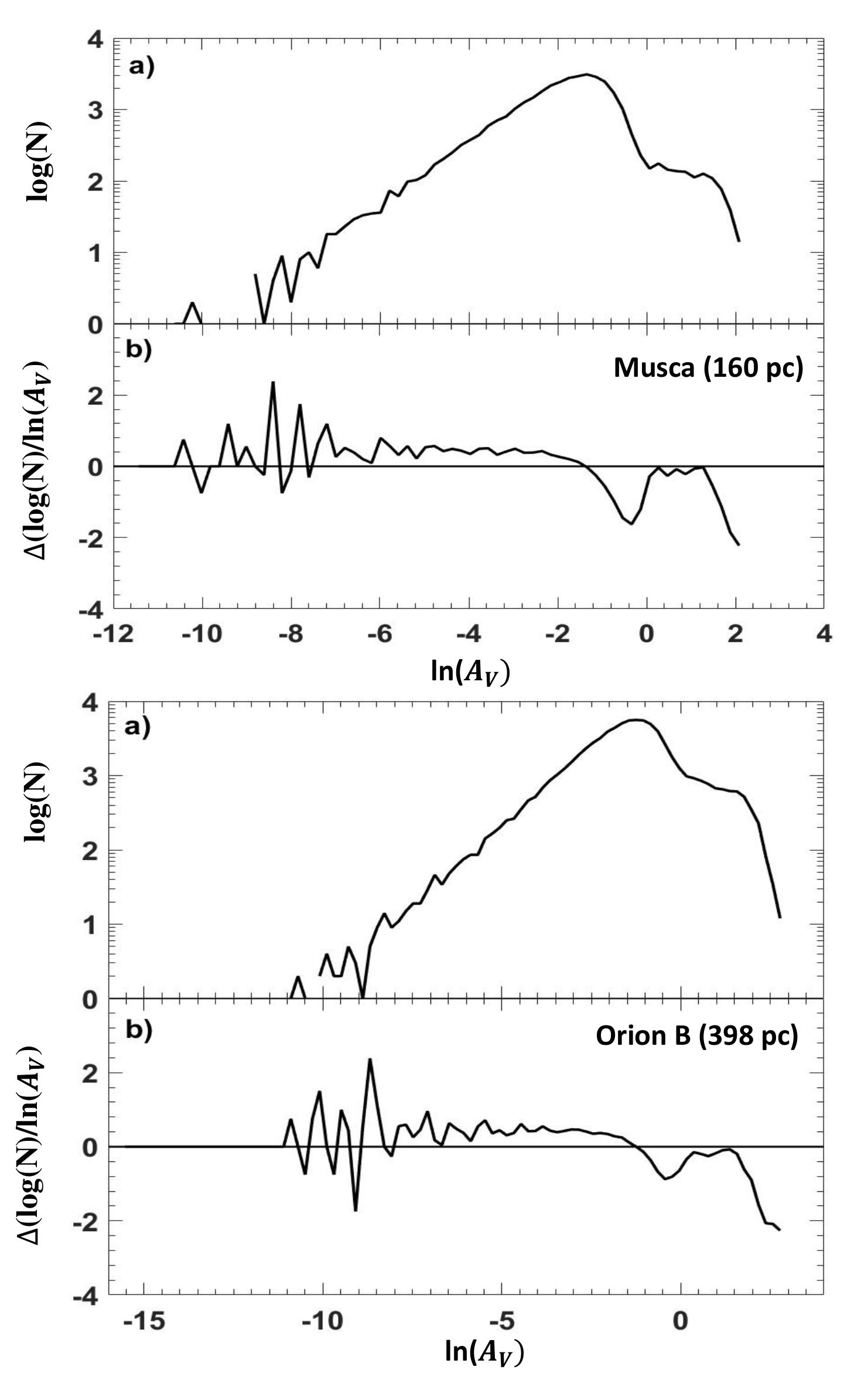}
    \caption{}
    \label{fig:FigureA3}
\end{figure}
\renewcommand{\thefigure}{\arabic{figure}}
\renewcommand{\thefigure}{\arabic{figure} (Cont.)}
\addtocounter{figure}{-1}
\begin{figure}
	\includegraphics[width=\columnwidth,height = 17cm]{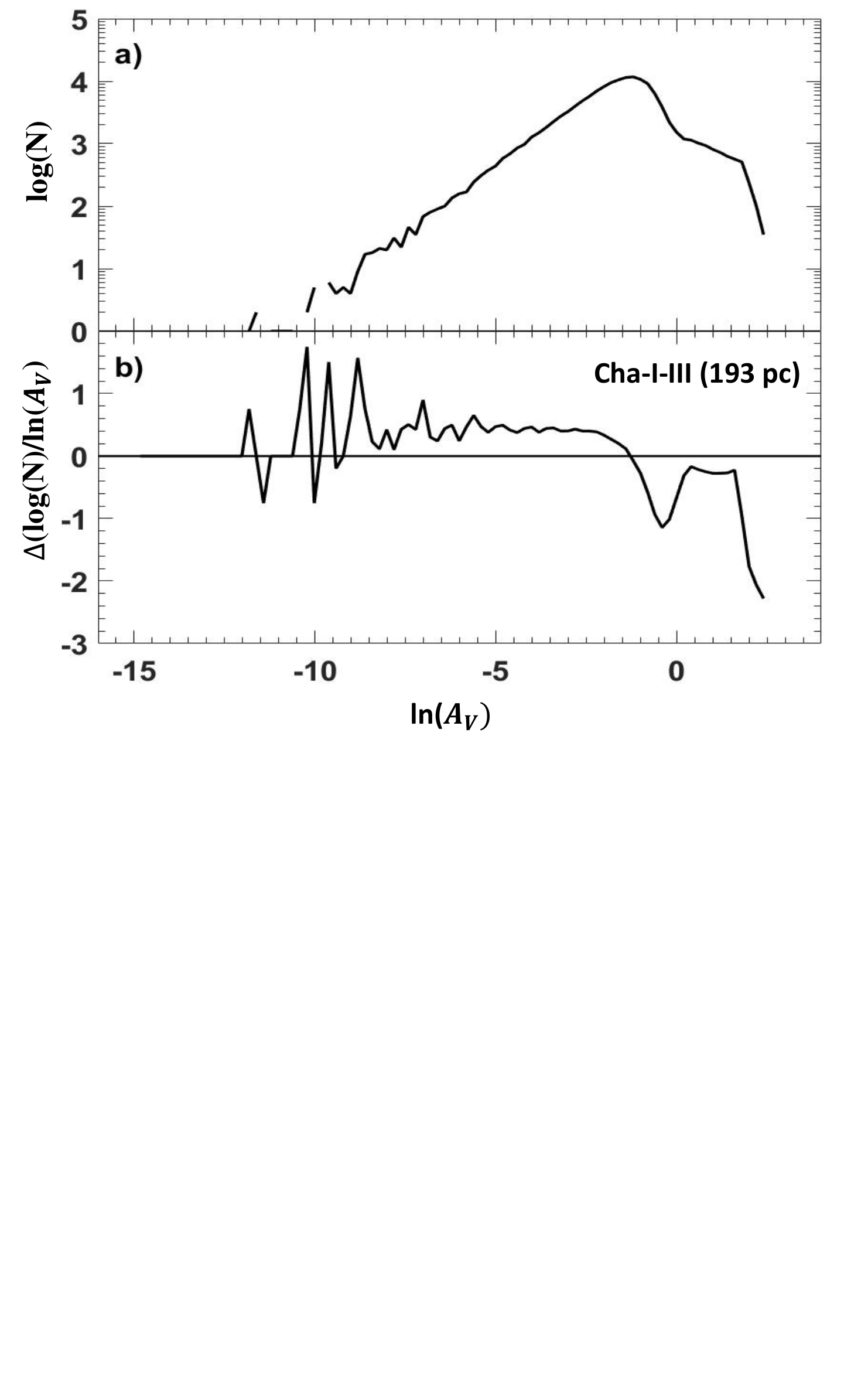}
    \caption{}
    \label{fig:FigureA4}
\end{figure}
\renewcommand{\thefigure}{\arabic{figure}}
\begin{figure}
	\includegraphics[width=\columnwidth,height = 17cm]{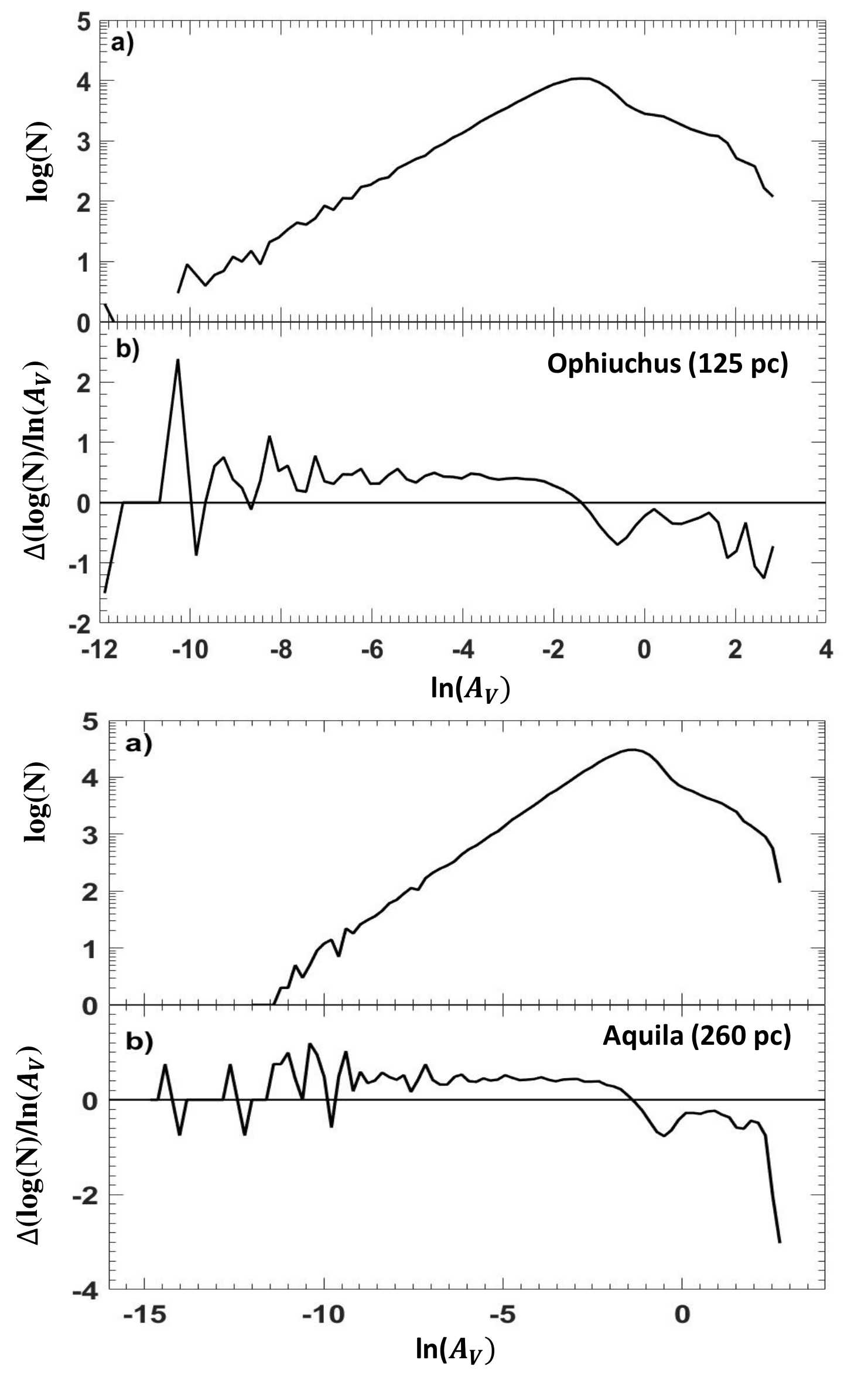}
    \caption{N-PDFs of molecular clouds regions as closer to be parallel with B-field by L17.}
    \label{fig:FigureA5}
\end{figure}
\renewcommand{\thefigure}{\arabic{figure} (Cont.)}
\addtocounter{figure}{-1}
\begin{figure}
	\includegraphics[width=\columnwidth,height = 17cm]{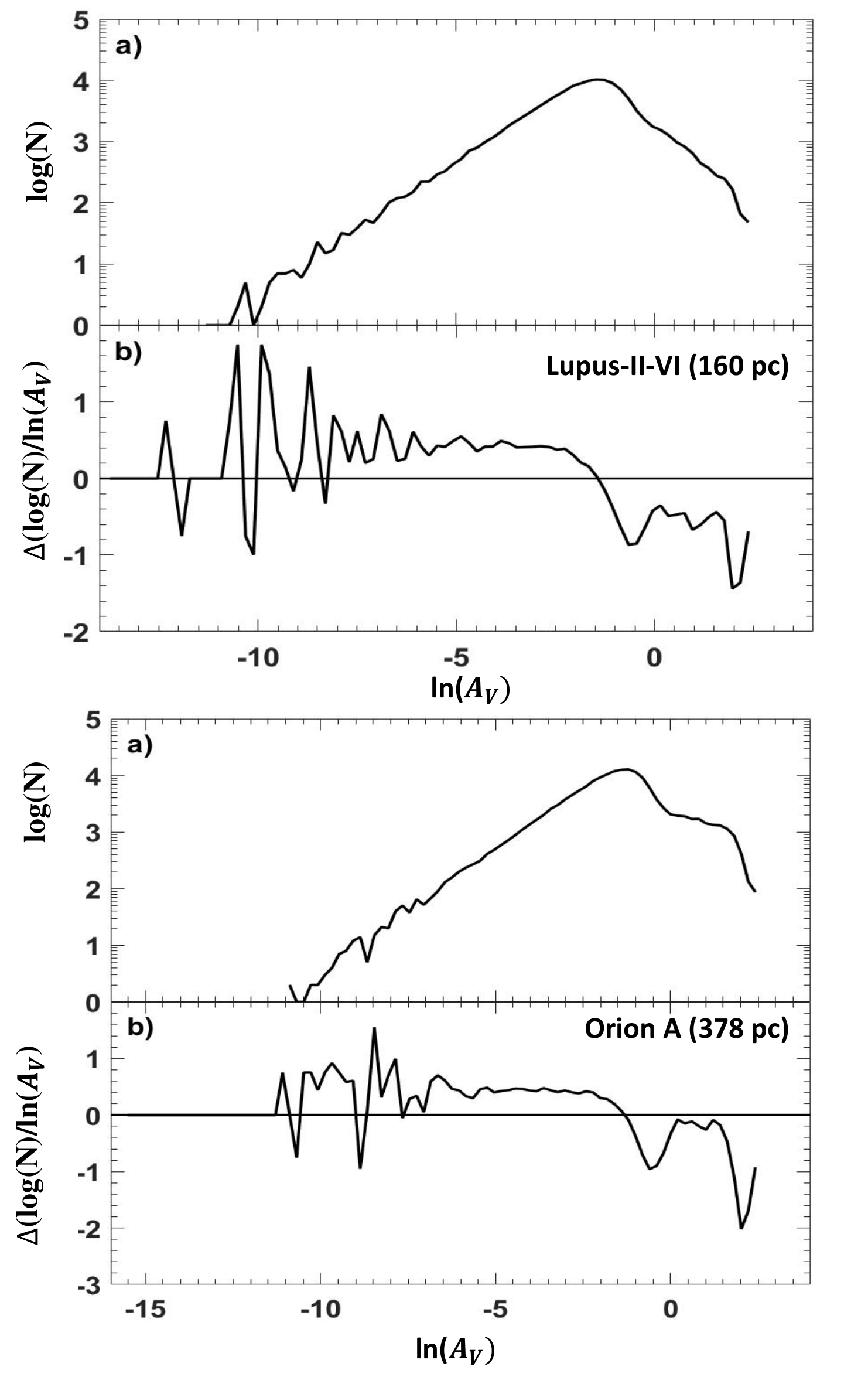}
    \caption{}
    \label{fig:FigureA6}
\end{figure}
\renewcommand{\thefigure}{\arabic{figure} (Cont.)}
\addtocounter{figure}{-1}
\begin{figure}
	\includegraphics[width=\columnwidth,height = 17cm]{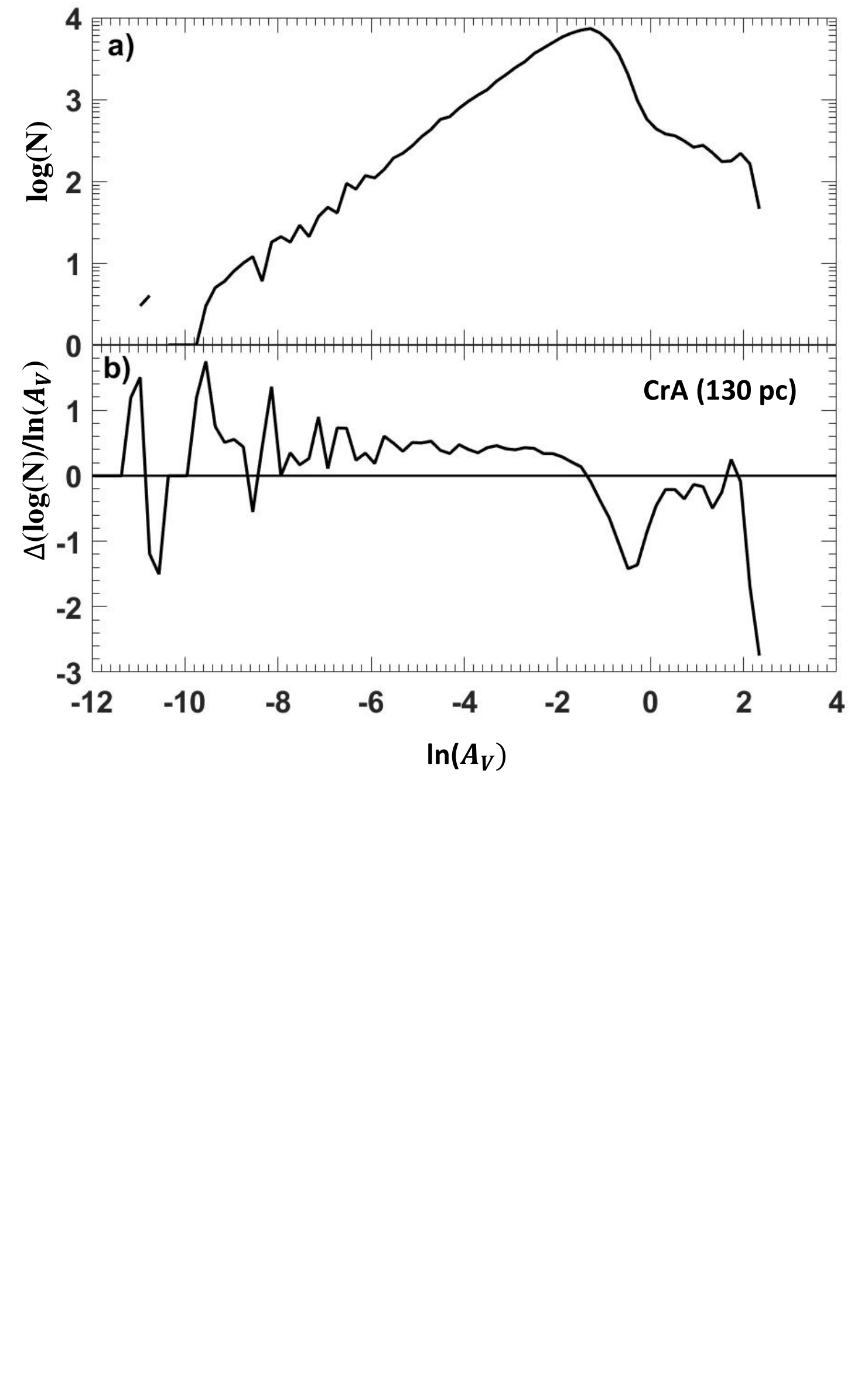}
   \caption{}
    \label{fig:FigureA7}
\end{figure}
\renewcommand{\thefigure}{\arabic{figure} (Cont.)}
\begin{table}
\raggedright
\caption{Transition density of the molecular clouds}
\begin{tabular}{|c|c|}
\hline
Cloud & Transition density \\ \hline
Musca & 0.71 \\ \hline
Corona & 0.62 \\ \hline
Lupus I & 0.73 \\ \hline
Lupus II-VI & 0.52 \\ \hline
Pipe & 0.49 \\ \hline
Chameleon & 0.67 \\ \hline
Ophiuchus & 0.55 \\ \hline
Taurus & 0.79 \\ \hline
IC5146 & 0.74\\ \hline
Aquila & 0.61 \\ \hline
Orion B & 0.64\\ \hline
Orion A & 0.55 \\ \hline
\end{tabular}
\begin{tablenotes}
      \small
      \item  The transition density of all molecular clouds in this work. 
    \end{tablenotes}
\end{table}


\bsp	
\label{lastpage}
\end{document}